\documentclass[a4paper, 11 pt]{article}
\pdfoutput = 1

\usepackage{jheppub, bm}
\pdfoutput=1
 \allowdisplaybreaks
\usepackage{graphicx}
\usepackage[dvipsnames]{xcolor}
\usepackage{amsmath}
\allowdisplaybreaks
\usepackage{amssymb}

\usepackage{slashed}
\usepackage[utf8]{inputenc}
\usepackage{color}
\usepackage[normalem]{ulem}
\usepackage{soul}
\usepackage{units}
\usepackage{rotating}
\usepackage{hhline,multirow,tabularx}
\usepackage{bm}
\usepackage{hyperref}
\usepackage{comment}
\usepackage[export]{adjustbox}
\usepackage{mdframed}

\def\bea#1\eea{\begin{align}#1\end{align}} 

\newcommand{\bef}{\begin{figure}[htb]\centering}
\newcommand{\eef}{\end{figure}}
\usepackage{mathtools}
\usepackage{tikz}
\usetikzlibrary{arrows.meta, decorations.pathreplacing, positioning, calc, decorations.pathmorphing}

\newcommand{\nn}{\nonumber}

\def\<{\langle}
\def\>{\rangle}

\def\d{\delta}  

\def\({\left(}
\def\[{\left[}
\def\){\right)}
\def\]{\right]}

\def\ln{\hbox{ln}}

\def\d{\mathrm d}

\usepackage{lipsum}

\begin{document}
\preprint{}

\makeatletter
\def\@fpheader{~}
\makeatother

\title{N$^{\mathbf{3}}$LL + $\mathcal{O}(\alpha_s^2)$ predictions of lepton-jet azimuthal angular distribution in deep-inelastic scattering}

\author[a]{Shen Fang,}

\author[a]{Mei-Sen Gao,}

\author[b]{Hai Tao Li}

\author[a,c,d]{and Ding Yu Shao}

\affiliation[a]{Department of Physics and Center for Field Theory and Particle Physics, Fudan University, Shanghai 200438, China}

\affiliation[b]{School of Physics, Shandong University, Jinan, Shandong 250100, China}

\affiliation[c]{Key Laboratory of Nuclear Physics and Ion-beam Application (MOE), Fudan University, Shanghai 200438, China}

\affiliation[d]{Shanghai Research Center for Theoretical Nuclear Physics, NSFC and Fudan University, Shanghai 200438, China}

\emailAdd{sfang23@m.fudan.edu.cn}
\emailAdd{msgao@fudan.edu.cn}
\emailAdd{haitao.li@sdu.edu.cn}
\emailAdd{dingyu.shao@cern.ch}

\abstract{We present an analysis of lepton-jet azimuthal decorrelation in deep-inelastic scattering (DIS) at next-to-next-to-next-to-leading logarithmic (N$^{3}$LL) accuracy, combined with fixed-order corrections at $\mathcal{O}(\alpha_s^2)$. In this study, jets are defined in the lab frame using the anti-$k_T$ clustering algorithm and the winner-take-all recombination scheme. The N$^{3}$LL resummation results are derived from the transverse-momentum dependent factorization formula within the soft-collinear effective theory, while the $\mathcal{O}(\alpha_s^2)$ fixed-order matching distribution is calculated using the {\tt NLOJET++} event generator. The azimuthal decorrelation between the jet and electron serves as a critical probe of the three-dimensional structure of the nucleon. Our numerical predictions provide a robust framework for precision studies of QCD and the nucleon's internal structure through jet observables in DIS. These results are particularly significant for analyses involving jets in HERA data and the forthcoming electron-ion collider experiments.}

\maketitle

\section{Introduction}

The primary goal of the future electron-ion collider (EIC) \cite{Accardi:2012qut, AbdulKhalek:2021gbh, Anderle:2021wcy, Abir:2023fpo} is to explore the structure of hadrons and nuclei, including their multi-dimensional partonic structure and gluon saturation. Recent investigations at both RHIC and the LHC have demonstrated that jets are effective tools for probing the inner structure of the nucleon \cite{Boer:2003tx, Vogelsang:2005cs,  STAR:2007yqh, Qiu:2007ey, Bomhof:2007su, Aschenauer:2016our, STAR:2017akg,  Chien:2019gyf,  Kang:2019ahe, Kang:2020xez, Kang:2020xyq, Gao:2023ulg}. Consequently, jet physics at the EIC is rapidly emerging as a significant area of research~\cite{ Gutierrez-Reyes:2018qez, Liu:2018trl,Gutierrez-Reyes:2019vbx, Liu:2020dct, Makris:2020ltr, Arratia:2020azl, Arratia:2020ssx,  Arratia:2020nxw, delCastillo:2020omr, Kang:2020fka, Kang:2020xgk, Kang:2021ryr, Kang:2021ffh, Kang:2021kpt, Zhang:2021tcc,  Hatta:2021jcd, Cirigliano:2021img,  delCastillo:2021znl, Liu:2021lan, Li:2021uww, Yan:2021htf, Yan:2022npz, Arratia:2022oxd, Burkert:2022hjz, Lee:2022kdn, Lai:2022aly, delCastillo:2023rng, Caucal:2023fsf,  Caucal:2023nci, Banfi:2023mhz, Tong:2023bus, Fang:2023thw, Buonocore:2024pdv, Borsa:2024rmh, Caucal:2024vbv}.

Precision calculations in deep-inelastic scattering (DIS) are essential for enhancing our understanding of partonic interactions and the internal structure of nucleons. Achieving high accuracy in these calculations requires the resummation of large logarithms to all orders and the inclusion of fixed-order corrections. Recent works have achieved next-to-next-to-leading-order (NNLO) accuracy in perturbative calculations of semi-inclusive deep-inelastic scattering (SIDIS) \cite{Goyal:2023zdi, Bonino:2024qbh, Goyal:2024tmo, Bonino:2024wgg}. Next-to-next-to-next-to-leading logarithmic (N$^{3}$LL) resummation results for the transverse momentum distribution of inclusive hadron production are also available \cite{Scimemi:2019cmh}. Additionally, the status of high-order calculations has reached a remarkable N$^3$LO accuracy for jet production in DIS \cite{Currie:2018fgr, Gehrmann:2018odt}. Several global event shape distributions in DIS, such as thrust \cite{Kang:2015swk}, energy-energy correlation \cite{Li:2020bub, Li:2021txc} and 1-jettiness \cite{Cao:2024ota}, are known at N$^{3}$LL+$\mathcal{O}(\alpha_s^2)$ accuracy. These calculations reduce theoretical uncertainties and improve the reliability of predictions for experimental observables, making them significant for the analysis of current and future DIS data.

In the electron-nucleon lab frame of DIS, the transverse momentum imbalance between the final-state electron and jet provides a direct measure of the incoming quark's transverse momentum \cite{Liu:2018trl, Liu:2020dct}. By examining this imbalance, or the azimuthal angle decorrelation between the lepton and the jet, one can extract direct information about the transverse momentum dependent (TMD) parton distribution functions (PDFs) \cite{Arratia:2020nxw, Arratia:2022oxd}. Recent measurements at HERA have demonstrated that the transverse momentum imbalance between the lepton and the jet in electron-proton collisions is consistent with theoretical predictions based on TMD factorization \cite{H1:2021wkz, ZEUS:2024mhu}. These findings establish a solid foundation for future jet studies of 3D imaging of the nucleon at the forthcoming EIC.

Lepton-jet associated production is insensitive to final-state hadron fragmentation and is complementary to the traditional SIDIS process. Notably, defining jets with the recoil-free recombination scheme significantly simplifies the TMD factorization and resummation formulas \cite{Banfi:2008qs, Gutierrez-Reyes:2019vbx, Chien:2020hzh, Chien:2022wiq, Fang:2023thw}, thereby enabling precision perturbative calculations. QCD resummation of the transverse momentum imbalance of lepton and jet in the Breit frame has been studied at N$^{3}$LL accuracy \cite{Gutierrez-Reyes:2018qez, Gutierrez-Reyes:2019vbx}, and the next-to-next-to-leading logarithmic (NNLL) resummation of the azimuthal angular distribution of lepton and jet in the lab frame was recently finished in~\cite{Fang:2023thw} for both $ep$ and $eA$ collisions. In addition to high-order resummation based on the leading power factorization formula, efforts toward the power corrections to jet observables have also been studied \cite{delCastillo:2023rng}.

In this paper, we present a prediction of the azimuthal angular distribution of the lepton and the leading jet at N$^{3}$LL + $\mathcal{O}(\alpha_s^2)$ accuracy, where final-state jets are defined in the lab frame using the anti-$k_T$ clustering algorithm \cite{Cacciari:2008gp}, and the winner-take-all (WTA) recombination scheme \cite{Bertolini:2013iqa, Larkoski:2014uqa}. We apply the TMD factorization formula \cite{Fang:2023thw} derived in soft-collinear effective theory (SCET)~\cite{Bauer:2000yr, Bauer:2001ct, Bauer:2001yt, Bauer:2002nz, Beneke:2002ph} to obtain the N$^{3}$LL resummation formula, and the $\mathcal{O}(\alpha_s^2)$ fixed-order matching distribution is obtained using the {\tt NLOJET++} event generator \cite{Nagy:2001xb, Nagy:2005gn}. It should be noted that all resummation ingredients are known, except for the two-loop constant term in the jet function. This constant was extracted numerically~\cite{Gutierrez-Reyes:2019vbx} from the {\tt Event2} generator~\cite{Catani:1996vz}, and preliminary numerical results are also presented in \cite{Brune:2022cgr, scet2023}. As an independent cross-check, we use {\tt NLOJET++} to generate $\mathcal{O}(\alpha_s^2)$ fixed-order predictions for the back-to-back dijet process in $e^+e^-$ collisions and determine the constant term by comparing its results with the singular distribution obtained from the fixed-order expansion of the resummation formula. The details are given in the appendix \ref{app:epem}.

The paper is structured as follows: We introduce the kinematic setup in section \ref{sec:kin}. Then, we discuss the factorization formula and gather all resummation ingredient functions in section \ref{sec:facres}. In section \ref{sec:foexp}, we provide the expression of $\mathcal{O}(\alpha_s^2)$ fixed-order expansion of the resummation formula. We establish our resummation and matching strategy, and present numerical results in section \ref{sec:num}. We conclude in section \ref{sec:conclusion}.

\section{Kinematics}\label{sec:kin}
We consider the process in $ep$ collisions as
\begin{align}\label{eq:pro}
    e(\ell) + p(P) \rightarrow e(\ell') + J(P_J) + X\,,
\end{align}
where $e(\ell)$ refers to the incident electron with momentum $\ell$, $p(P)$ denotes the proton with momentum $P$, and $e(\ell')$ symbolizes the scattered electron with momentum $\ell'$. The final state includes a leading jet $J$ with momentum $P_J$ and inclusive particles $X$. In the lab frame, the four-momenta of the initial-state lepton and proton can be expressed in terms of the lepton energy $E_e$ and the proton energy $E_p$ as:
\begin{align}
    P^\mu = E_p \, n^\mu\,,
    \qquad
    \ell^\mu = E_e \, \bar{n}^\mu\,,
\end{align}
where $n^\mu=(1,0,0,1)$ and $\bar{n}^\mu=(1,0,0,-1)$ are the light-cone reference vectors in Cartesian space-time coordinates. In taking this parametrization, the $z$ direction is defined along the proton beam while the electron beam is in the negative $z$ direction.

At the leading order (LO), the partonic process is denoted by $e(l)+q(k)\to e(l')+q(P_J)$, where the final-state momenta of the lepton and the jet are parameterized in terms of the lepton's rapidity $y_l$ and the jet's rapidity $y_J$ and the transverse momentum $p_T$ as 
\begin{align}\label{eq:final}
    {\ell'}^\mu = p_T\left( \mathrm{ch}y_l,0,1,\mathrm{sh}y_l\right)\,,
\qquad
    P_J^\mu = p_T \left( \mathrm{ch}y_J,0,-1,\mathrm{sh}y_J\right)\equiv E_J \, n_J^\mu\,,
\end{align}
where $E_J = p_T \, \mathrm{ch}y_J$ represents the energy of the jet, and $n_J^\mu=\left( 1,0,-\mathrm{sech}y_J,\mathrm{th}y_J\right)$ denotes the light-cone vector along the jet direction. We can also define another jet light-cone vector $ {\bar n}^\mu_J = \left( 1,0,\mathrm{sech}y_J,-\mathrm{th}y_J\right).$ By momentum conservation, the four-momentum of the intermediate photon is given by $q = \ell-\ell'$. From this parametrization, we can define the following kinematic variables
\begin{align}
    & Q^2 = -(\ell-\ell')^2 = 2e^{-y_J} x_{\tt bj}\, E_p \, p_T \,, 
    \qquad
    x_{\tt bj} = \frac{Q^2}{2 P\cdot q} = \frac{E_e \, p_T \, e^{2y_J}}{E_e \, E_p \, 2\,e^{y_J}-E_p \, p_T }\,, \notag
    \\
    & 
    \hat{t} = -Q^2\,, 
    \qquad 
    \hat{s} = x_{\tt bj} \, S\,,
    \qquad
    \hat{u} = -\hat{s}-\hat{t}\,,
    \qquad
    S=4E_e \, E_p \,,
\end{align}
where we neglect the proton mass and $k^\mu=x_{\tt bj} P^\mu$. Based on the above definition, the LO cross section is expressed as
\begin{align}
    \frac{\d\sigma}{\d^2 p_T\, \d y_J} = \sigma_0 \sum_q e_q^2 f_{q/p}(x_{\tt bj}, \mu),
\end{align}
where $e_q$ is the fractional charge carried by parton $q$, $f_{q/p}$ denotes the collinear PDF and the prefactor $\sigma_0$ is given by 
\begin{align}
    \sigma_0 = \frac{\alpha_{\rm em}^2}{
     S Q^2}\frac{2\left(\hat{s}^2+\hat{u}^2\right)}{\hat{t}\hat{u}}\,.
\end{align}
The azimuthal angle between the final electron and the leading jet is defined as $\delta\phi\equiv|\pi-|\phi_J-\phi_l||$ with $-\pi<\phi_{l,J}\leq\pi$. At LO, $\delta\phi=0$, as given in Eq.~\eqref{eq:final}. Beyond the Born configuration, higher-order corrections will contribute.  In the back-to-back limit, $\delta\phi \ll 1$, one must consider the presence of any number of soft and collinear emissions, which causes a small deviation from $\delta\phi=0$. These contributions are captured by all-order resummation. Additionally, perturbative matching corrections need to be included as $\delta\phi\sim\mathcal{O}(1)$. In this work, we incorporate these corrections up to $\mathcal{O}(\alpha_s^2)$.

\section{Factorization and resummation} \label{sec:facres}

In the back-to-back limit, the TMD factorization formula for the azimuthal decorrelation has been derived for $V$+jet production in $pp$ collisions \cite{Chien:2020hzh, Chien:2022wiq} and $e$+jet production in $ep$ collisions \cite{Fang:2023thw} within the SCET framework. In this section, we will briefly review the result.  

The azimuthal decorrelation between the lepton ($l'$) and jet ($P_J$) represents the tangential offset $\lambda_x$, expressed as
\begin{align}
    \lambda_x \equiv l'_x + P_{x,J}.
\end{align}
In the limit $|\lambda_x|\ll Q$ with the jet radius $R\sim \mathcal{O}(1)$, the factorization theorem in $b$-space reads \cite{Fang:2023thw}
\begin{align}\label{eq:ep_fac}
    \frac{\d\sigma}{\d^2 p_T\, \d y_J\, \d \lambda_x} & = \,
     \sigma_0 \, H(Q,\mu)\! \int_{-\infty}^{+\infty} \frac{\d b_x}{2\pi} e^{i b_x \lambda_x} \sum_{q} e_{q}^{2} ~ {\cal B}_{q/p}\left(x_{\tt bj},b_x,\mu,\zeta_{\cal B}/\nu^2\right)\nn\\
&\times{\cal J}_{q}\left(b_x,\mu,\zeta_{\cal J}/\nu^2\right)\, {\cal S}\left(b_x,n\cdot n_J,\mu,\nu\right), 
\end{align}
where all ingredients, including the hard function $H$, TMD beam function $\mathcal{B}_{q/p}$, jet function $\mathcal{J}_{q}$ and soft function $\mathcal{S}$, depend on the renormalization scale $\mu$, and $\mathcal{B}_{q/p}$, $\mathcal{J}_{q}$ and $\mathcal{S}$ also depend on the rapidity scale $\nu$. Here $\zeta_\mathcal{B}=(\bar n \cdot k)^2$ and $\zeta_\mathcal{J}=(\bar n_J \cdot P_J)^2$ denote the Collins-Soper parameters of the beam and jet functions. The natural scales of the ingredients in Eq.~\eqref{eq:ep_fac} are 
\begin{align}
& \mu_h \sim \nu_{\mathcal{B}} \sim \nu_{\mathcal{J}} \sim Q \sim p_T, \notag\\
& \mu_\mathcal{B} \sim \mu_\mathcal{J} \sim \mu_\mathcal{S} \sim \nu_\mathcal{S} \sim |\lambda_x| \sim 1 /\left|b_x\right| . \notag
\end{align}
Explicitly, the components of the factorized formula \eqref{eq:ep_fac} are:
\begin{itemize}
    \item $H(Q,\mu)$ is the hard function taking into account virtual corrections at the scale $Q$, and it can be determined order-by-order in perturbation theory by a matching calculation in QCD and in SCET at the hard scale $\mu_h$. Its two-loop expression and corresponding anomalous dimensions needed at N$^3$LL accuracy have been given in the appendix \ref{app:adim}. 
    
    \item The unsubtracted quark TMD PDF $\mathcal{B}_{q/p}$ is the same one that appears in the SIDIS factorization, describing the transverse momentum of the colliding hard parton with respect to the beam axis due to collinear initial-state radiation. In the small $b$ limit, it can be perturbatively matched onto the collinear PDFs via the operator product expansion as follows
    \begin{align}\label{eq:OPE}
    \mathcal{B}_{q/p}\left(x,b_x,\mu,\zeta_\mathcal{B}/\nu^2\right) &=\sum_i \int_{x}^1 \frac{\d {z }}{z} \mathcal{I}_{q / i}(z, b_x, \mu, \zeta_{\mathcal{B}}/\nu^2 ) f_{i/p}(x/z, \mu)\\
    &\equiv\sum_i\mathcal{I}_{q/i}\left(x,b_x,\mu,\zeta_\mathcal{B}/\nu^2\right)
    \otimes f_{i/p}(x, \mu)\,,\nn
    \end{align}
    where $f_{i/p}$ denotes the collinear PDF. The expressions of the matching coefficient $\mathcal{I}_{q / i}$ are known in \cite{Catani:2011kr, Catani:2012qa, Gehrmann:2012ze, Gehrmann:2014yya, Luebbert:2016itl, Echevarria:2016scs, Luo:2019hmp, Luo:2019bmw} at two-loop and \cite{Behring:2019quf, Luo:2019szz, Luo:2020epw, Ebert:2020yqt} at three-loop. At large $b$ non-perturbative contribution becomes significant, and we include the non-perturbative models~\cite{Sun:2014dqm, Kang:2015msa, Echevarria:2020hpy} in the following numerical calculation. 

    \item The jet function $\mathcal{J}_{q}$ takes into account the collinear dynamics of the jet formation, which describes the offset of the WTA axis with respect to the jet momentum. The operator definition can be found in \cite{Gutierrez-Reyes:2018qez, Gutierrez-Reyes:2019vbx}. It is noted that there is no transverse momentum dependence for the jet axis defined with the $E$-scheme in the narrow cone limit \cite{Liu:2018trl, Chien:2019gyf, Liu:2020dct}, since it is a non-global observable \cite{Dasgupta:2001sh}. In this work, as the direction of the jet axis with the WTA scheme is insensitive to soft emissions, the use of this axis removes the non-global logarithms (NGLs) \cite{Banfi:2008qs, Chien:2020hzh, Chien:2022wiq} and thus avoids the complications associated with resumming NGLs \cite{Becher:2015hka, Caron-Huot:2015bja, Larkoski:2015zka, Becher:2016mmh, Larkoski:2016zzc, Hatta:2017fwr, Balsiger:2019tne, Banfi:2021owj, Banfi:2021xzn, Becher:2023vrh}.

    \item The soft function $\mathcal{S}$ in the factorization formula \eqref{eq:ep_fac} differs from the standard TMD soft function $\mathcal{S}_\perp$ in SIDIS factorization \cite{Boussarie:2023izj}. However, it can be determined from $\mathcal{S}_\perp$ by performing a boost, especially, since the observable $\lambda_x$ is perpendicular to the boost. Only the rapidity regulator is affected \cite{Kasemets:2015uus, Gao:2019ojf, Gao:2023ivm}, yielding
    \begin{align}
        \mathcal{S}\left(b_x, n\cdot n_J, \mu, \nu\right)= \mathcal{S}_\perp(b_x, \mu, \nu \sqrt{n \cdot n_J / 2}).
    \end{align}
\end{itemize}

Similar to Ref. \cite{Fang:2023thw}, we apply the standard CSS treatment \cite{Collins:2011zzd} to perform the rapidity logarithm resummation. We define properly subtracted TMD PDFs and jet functions, independent of the rapidity scale $\nu$, as follows
\begin{align}
f_{q/p}\left(x_{\tt bj},b_x,\mu,\zeta_B\right) & =  \mathcal{B}_{q/p}\left(x_{\tt bj},b_x,\mu,\zeta_\mathcal{B}/\nu^2\right)  \sqrt{\mathcal{S}_{\perp}\left(b_x,\mu,\nu\right)}\,, \\
J_q\left(b_x,\mu,\zeta_J\right) &= \mathcal{J}_q\left(b_x,\mu,\zeta_\mathcal{J}/\nu^2\right)\, \frac{\mathcal{S}\left(b_x,n\cdot n_J,\mu,\nu\right)}{\sqrt{\mathcal{S}_{\perp}\left(b_x,\mu,\nu\right)}}\,,
\end{align}
where we have redefined the Collins-Soper scales in the conventional manner  as 
\begin{align}
    \sqrt{\zeta_B} = \sqrt{\zeta_\mathcal{B}}\,,
    \qquad
    \sqrt{\zeta_J} = \frac{n\cdot n_J}{2}\sqrt{\zeta_\mathcal{J}}\,,
\end{align}
with $\zeta_B \, \zeta_J = Q^4$. 

Finally, we obtain the generic solutions of the TMD PDF and jet function by performing the integration over the scales $\mu$ and $\zeta_{B,J}$. The all-order resummation is achieved by evolving the different components in Eq.~(\ref{eq:ep_fac}) from their natural scales to the scales $\{\mu_h,\zeta_f\}$. The resulting formula is expressed as
\begin{align}\label{eq:res}
    & \frac{\d\sigma}{\d^2 p_T\, \d y_J\, \d\delta\phi} = \sigma_0  p_T H\left(Q,\mu_h\right)\!\!\int_{0}^\infty \frac{2\d b}{\pi} e^{i b p_T \delta\phi}
    \sum_q e_q^2\, f_{q/p}(x_{\tt bj},b,\mu_h,\zeta_f) J_q(b,\mu_h,\zeta_f) \,,
\end{align}
where we define $b\equiv|b_x|$ and take $\delta\phi=|\lambda_x|/p_T$ in the back-to-back limit. The TMD PDF and jet function at scales $\{\mu_h,\zeta_f\}$ read 
\begin{align}\label{eq:TMDPDF}
    f_{q/p}  \left(x_{\tt bj},b,\mu_h,\zeta_f\right)  & = f_{q/p}\left(x_{\tt bj},b,\mu_b,\zeta_{i}\right) 
 \exp\left[\int_{\mu_b}^{\mu_h} \frac{\d\mu}{\mu}\gamma^f_\mu\left(\mu,\zeta_f\right)\right]
\left(\frac{\zeta_f}{\zeta_{i}}\right)^{\frac{1}{2}\gamma^f_\zeta(b,\,\mu_b)}\,,
\end{align}
and
\begin{align}\label{eq:TMDjet}
J_{q}  \left(b,\mu_h,\zeta_f\right) &= J_{q}\left(b,\mu_b,\zeta_{i}\right) \exp\left[\int_{\mu_b}^{\mu_h} \frac{\d\mu}{\mu}\gamma^J_\mu\left(\mu,\zeta_f\right)\right]
\left(\frac{\zeta_f}{\zeta_{i}}\right)^{\frac{1}{2}\gamma^J_\zeta\left(b,\,\mu_b\right)} \,,
\end{align}
respectively, where anomalous dimensions are given by
\begin{align}\label{eq:AD_jet}
\gamma^{f/J}_\mu\left(\mu,\zeta_f\right)&=C_F\gamma_{\rm cusp}(\alpha_s)\ln{\frac{\mu^2}{\zeta_f}}+\gamma_{f/J}(\alpha_s)\,,\\
\gamma^{f/J}_\zeta(b,\,\mu_b) & = \int_{\mu_b^2}^{b_0^2/b^2} \frac{\d\mu^2}{\mu^2}C_F\gamma_{\rm cusp}(\alpha_s)+\gamma_r(\alpha_s)\,,\nn
\end{align}
with $b_0 \equiv 2 e^{-\gamma_E}$. All the necessary anomalous dimensions at N$^3$LL are given in the appendix \ref{app:adim}. Notably, unlike the resummation formula presented in \cite{Liu:2018trl, Liu:2020dct}, Eq.~\eqref{eq:res} no longer depends on the remainder soft factor due to the jet definition in the WTA recombination scheme. The cross section is simply expressed as the product of the hard function, TMD PDF, and jet function.

\section{Fixed-order expansion}\label{sec:foexp}

In this section, we provide the fixed-order singular cross section at the level of the integrated cross section, which is defined as
\begin{align}
    \frac{\d}{\d^2 p_T\, \d y_J}\sigma_{\mathrm{singular}}(\delta\phi^{\mathrm{cut}})\equiv\int^{\delta\phi^{\mathrm{cut}}}_{0}
    \d\delta\phi \frac{\d\sigma}{\d^2 p_T\, \d y_J\, \d\delta\phi}\,. 
\end{align}
Here the singular cross section can be recovered from the resummation formula \eqref{eq:res} by setting $\mu_b=\mu_h$ and $\zeta_{f} = \zeta_{i}=Q^2$, which is expressed as
\begin{align}
   \frac{\d}{\d^2 p_T\, \d y_J} \sigma_{\mathrm{singular}}(\delta\phi^{\mathrm{cut}})=\sigma_0\,  \sum_q e_q^2\sum_i \int_{x_{\tt bj}}^1 \frac{\d z}{z} C_{q/i}\left(z,\delta\phi^{\mathrm{cut}},Q\right)  f_{i/p}\left(x_{\tt bj}/z,Q\right)\,, 
\end{align}
where we choose $\mu_h=Q$. The perturbative coefficients of $C_{q/i}$ have the following expansion in the strong coupling constant
\begin{align}   C_{q/i}\left(z,\delta\phi^{\mathrm{cut}},Q\right) = \sum_{n=0}^\infty C_{q/i}^{(n)}\left(z,\delta\phi^{\mathrm{cut}},Q\right)\left(\frac{\alpha_s(Q)}{4\pi}\right)^{n}\,. 
\end{align}
Writing the perturbative coefficient $C_{q/i}^{(n)}$ as a function of the logarithm $L\!\equiv\mathrm{ln}\!\left(2 p_T\delta\phi^{\mathrm{cut}}/Q\right)$, up to $\mathcal{O}(\alpha_s^2)$, we have
\begin{align}\label{eq:singular}
C_{q/i}^{(0)}\left(z,\delta\phi^{\mathrm{cut}},Q\right)&=\delta_{qi}\delta(1-z),\\
C_{q/i}^{(1)}\left(z,\delta\phi^{\mathrm{cut}},Q\right)&=\delta_{qi}\delta(1-z)C_F\left(-8L^2-12L
-9-\frac{4\pi^2}{3}-6\mathrm{ln}2\right)+2P_{qi}^{(0)}(z)L+I_{qi}^{(1)}(z),\notag\\
C_{q/i}^{(2)}\left(z,\delta\phi^{\mathrm{cut}},Q\right)&=\delta_{qi}\delta(1-z)\left(C_F^2 L_F+C_F C_A L_A+C_F T_F n_f L_f +j_{2}
    \right)+ A_{qi}^{(2)}(z)+I_{qi}^{(2)}(z),\nn
\end{align}
with 
\begin{align}
L_F&=32L^4 + 96L^3 + L^2\left( 144+\frac{64\pi^2}{3}+48\mathrm{ln}2 \right) 
+ L\left( 102+40\pi^2+72\mathrm{ln}2+160\zeta_3 \right)\nn\\ 
&+\frac{63}{4} + \frac{67 \pi^2}{2} + \frac{851 \pi^4}{360} + 96 \mathrm{ln}2 + 3 \pi^2 \mathrm{ln}2 + 132 \zeta_3\,, \notag\\
L_A&=\frac{352}{9}L^3+L^2\left(-\frac{140}{9}+\frac{8\pi^2}{3}\right)
+L\left(-\frac{188}{3}+\frac{88\pi^2}{9}+44\mathrm{ln}2+48\zeta_3\right)\\
&-\frac{46301}{324} - \frac{1015 \pi^2}{108} + \frac{23\pi^4}{30} + \frac{1253}{9} \zeta_3
\,, \notag\\
L_f&=-\frac{128}{9}L^3+\frac{16}{9}L^2+
L\left(\frac{64}{3}-\frac{32\pi^2}{9}-16\mathrm{ln}2\right)
+\frac{3757}{81} +\frac{65 \pi^2}{27} - \frac{220}{9}\zeta_3\,.  \nn
\end{align}
Here, $j_2$ denotes the two-loop constant term of the jet function, which has been numerically determined using {\tt Event2}~\cite{Gutierrez-Reyes:2019vbx}, and some preliminary numerical results were also presented in \cite{Brune:2022cgr,scet2023}. For independent cross-checks, we use {\tt NLOJET++} to determine the constant term. Details are given in appendix \ref{app:epem}. The one- and two-loop splitting functions $P_{ji}^{(n)}$ and scale-independent coefficients $I_{qi}^{(n)}$ are given in Eqs.~\eqref{splitting} and \eqref{eq:matching_coefficient}, respectively, and 
\begin{align}\label{eq:A2}
A_{qi}^{(2)}(z)&=\left[-8C_FL^2+L\left(-\frac{22}{3}C_A+\frac{8}{3}T_F n_f-12C_F\right) +C_F\left(
-9-\frac{4\pi^2}{3}-6\mathrm{ln}2\right)\right]I_{qi}^{(1)}(z)\notag\\
&+ \left[-16C_F L^3
+L^2\left(-\frac{44}{3}C_A+\frac{16}{3}T_F n_f-24C_F\right)
+C_F L\left(-18-\frac{16\pi^2}{3}-12\mathrm{ln}2\right)
\right.\notag \\
& \left.-\frac{11\pi^2}{9}C_A+\frac{4\pi^2}{9}T_F n_f+C_F\left(-2\pi^2-32\zeta_3\right)\right]P_{qi}^{(0)}(z)+2LP_{qi}^{(1)}(z)\notag \\
&+2L\sum_j I_{qj}^{(1)}(z) \otimes P_{ji}^{(0)}(z)
+\left(2L^2+\frac{\pi^2}{6}\right)\sum_j P_{qj}^{(0)}(z) \otimes P_{ji}^{(0)}(z) \,.
\end{align}

\section{Numerical results} \label{sec:num}

In this section, we present the numerical results for both EIC and HERA. The kinematic cuts applied are as follows:
\begin{itemize}
\item EIC: $E_e = 18~{\rm GeV}$, $E_p = 275~{\rm GeV}$, $p_T > 15~{\rm GeV}$, $|y_J| < 1$,
\item HERA: $E_e = 27.5~{\rm GeV}$, $E_p = 920~{\rm GeV}$, $p_T > 10~{\rm GeV}$, $Q^2 > 150~{\rm GeV}^2$, $-1 < y_J < 2.5$.
\end{itemize}
Jets are reconstructed using the anti-$k_T$ algorithm \cite{Cacciari:2008gp} and the WTA recombination scheme \cite{Bertolini:2013iqa, Larkoski:2014uqa} with $R=1$. 

\begin{figure}[t]
    \centering
    \includegraphics[scale=0.5]{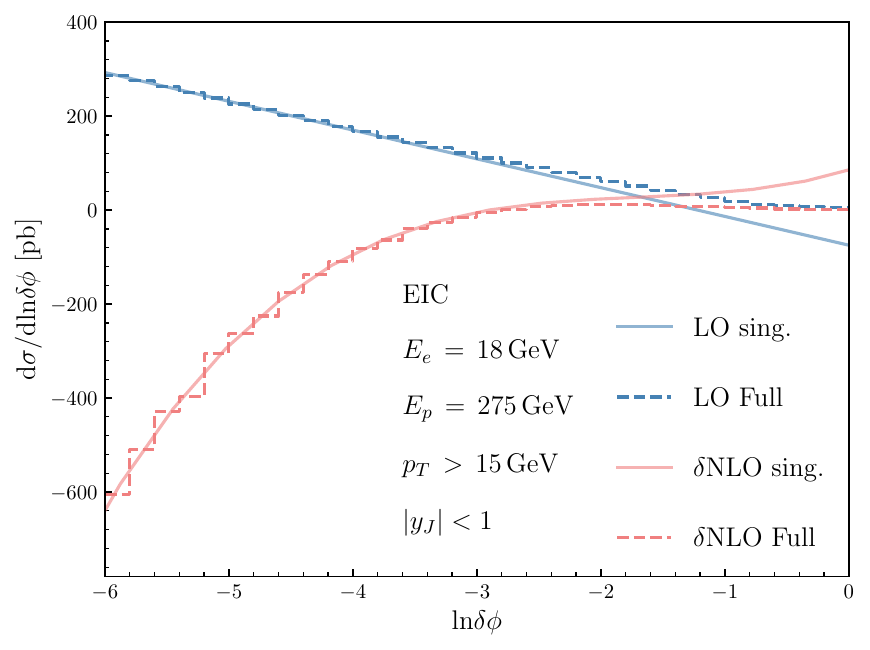}
    \includegraphics[scale=0.5]{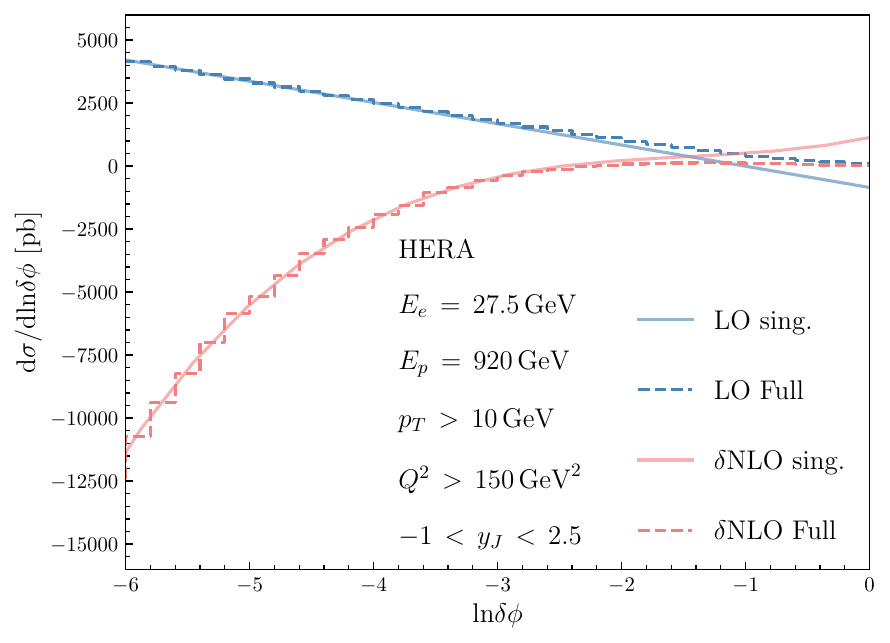}
    \caption{  $\delta\phi$ differential distributions at the EIC (left) and HERA (right). Solid lines represent LO (blue) and $\delta$NLO (red) singular distributions, while dashed lines represent LO (blue) and $\delta$NLO (red) fixed-order results obtained from {\tt NLOJET++}.
    }
    \label{fig:FO-expansion}
\end{figure}

Figure \ref{fig:FO-expansion} illustrates the singular and fixed-order $\delta\phi$ differential distributions for the EIC (left) and HERA (right). The solid lines denote LO (blue) and $\delta$NLO (red) singular distributions, where LO and $\delta$NLO denote the contribution from $C^{(1)}_{q/i}$ and $C^{(2)}_{q/i}$ in Eq.~\eqref{eq:singular}, respectively. The dashed lines represent LO (blue) and $\delta$NLO (red) fixed-order results obtained from {\tt NLOJET++} \cite{Nagy:2001xb, Nagy:2005gn} and {\tt FASTJET} \cite{Cacciari:2011ma}. At small values of $\ln \delta\phi$, the LO and $\delta$NLO singular terms (solid lines) and the full results (dashed lines) are consistent, indicating the dominance of singular terms in this region. This observation confirms the correct inclusion of large logarithmic terms in our factorization theorem at $\mathcal{O}(\alpha_s^2)$. As $\ln \delta\phi$ increases, discrepancies between the solid and dashed lines become evident, highlighting the increasing significance of matching corrections to the distribution. These matching corrections are absent in the resummation formula and can be included by incorporating the fixed-order calculations.

\begin{figure}[t]
    \centering
    \includegraphics[scale=0.45]{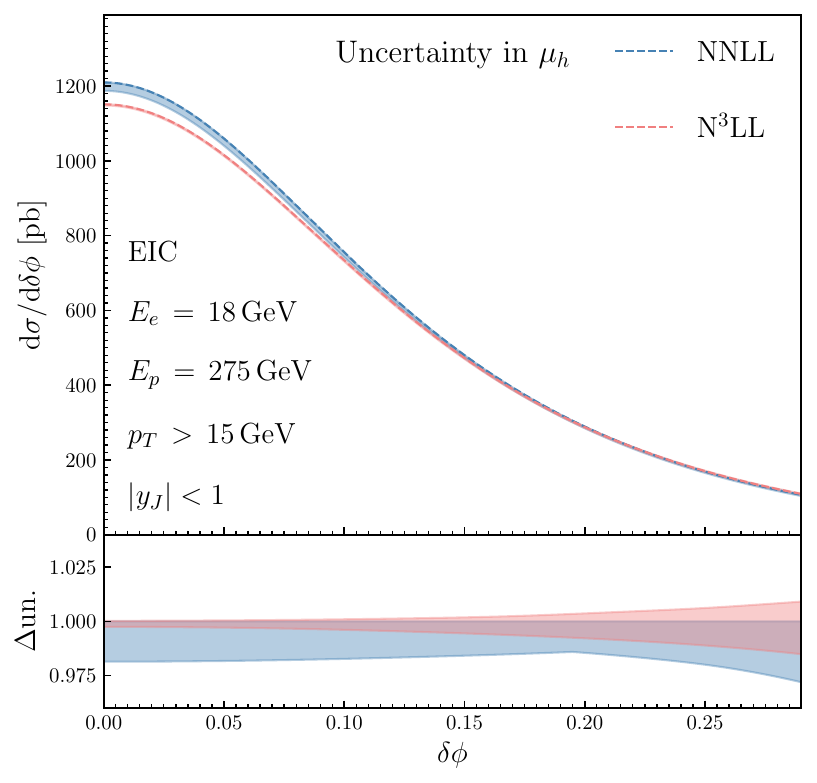}~
    \includegraphics[scale=0.45]{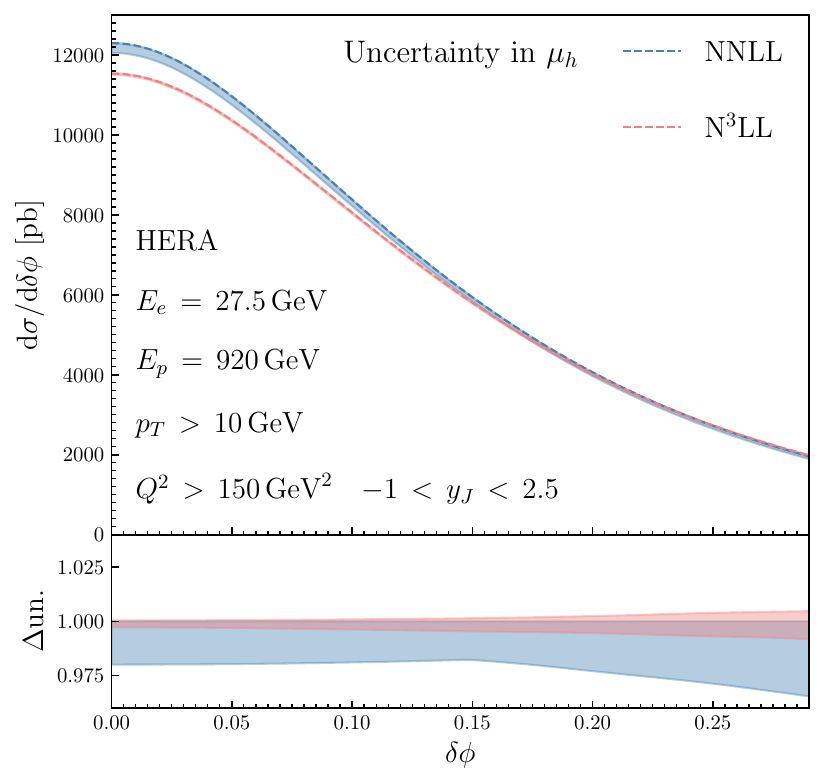} \\
    \includegraphics[scale=0.45]{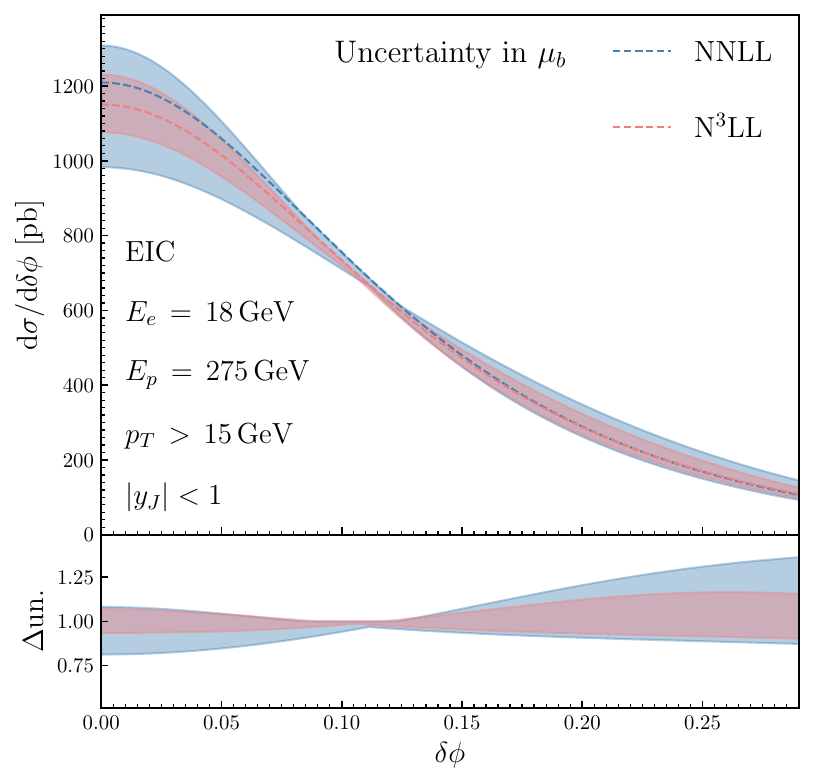}~
    \includegraphics[scale=0.45]{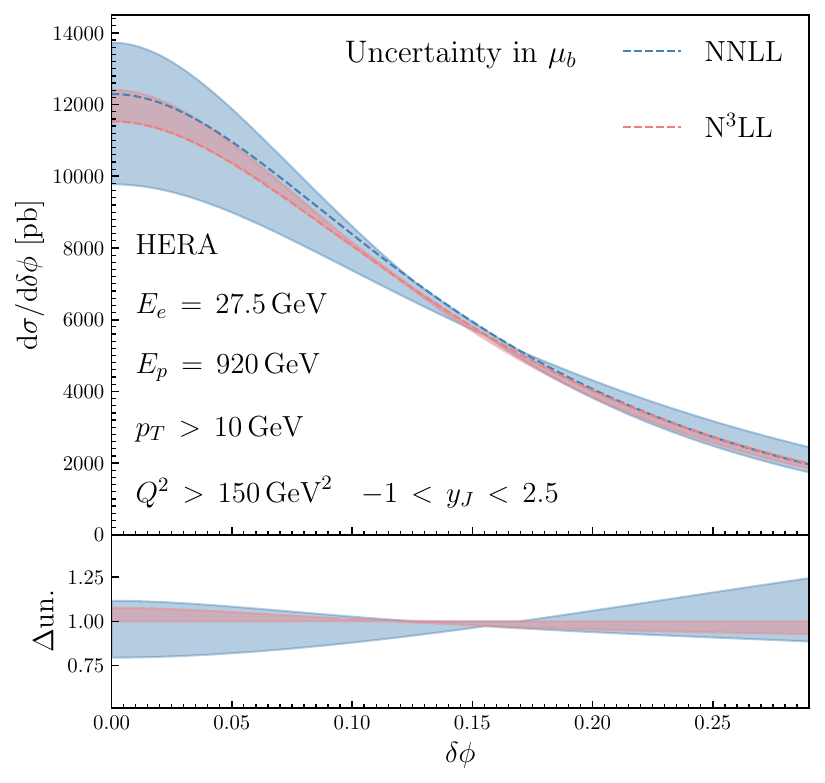}
    \caption{ Comparison of resummation results at NNLL (blue) and N$^3$LL (red) for the EIC (left panels) and HERA (right panels). The uncertainty estimates are shown for variations in the renormalization scales $\mu_h$ (upper panels) and $\mu_b$ (lower panels), where each scale is varied up and down by a factor of two. 
    }
    \label{fig:res}
\end{figure}

Below, we present the results of the resummation calculations using  Eq.~\eqref{eq:res}. In this paper, the natural scales for the resummation ingredients in Eq.~\eqref{eq:res} are chosen as follows:
\begin{align}\label{eq:resscales}
    \mu_h = Q, \quad \mu_b = b_0/b_*, \quad \zeta_{i} = b_0^2/b^2,\quad \zeta_f = Q^2\,.
\end{align}
 To avoid the Landau pole at the scale $\mu_b\sim 1/b$ as $b\to\infty$, we have applied the $b_*$-prescription \cite{Collins:1984kg}, where $b_*$ is defined by
\begin{align}
    b_* \equiv b/\sqrt{1+b^2/b_{\rm max}^2}\,,
\end{align}
with $b_{\rm max}=1.5$ GeV$^{-1}$. Besides, in the large $b$ region, we also include the non-perturbative Sudakov factor $U_{\rm NP}$ in TMD PDFs, given by \cite{Sun:2014dqm,Kang:2015msa,Echevarria:2020hpy,Alrashed:2021csd}
\begin{align}
    U_{\rm NP}(b,Q_0,Q) = \exp\left[-g_1 b^2-\frac{g_2}{2}  \ln \frac{Q}{Q_0} \ln \frac{b}{b_*}\right]\,, 
\end{align}
with $g_1=0.106\,\mathrm{GeV}^{2}$, $g_2=0.84$ and $Q_0^2=2.4\, \mathrm{GeV}^2$. Consequently, the TMD PDF is modified to
\begin{align}
    f_{q/p}  \left(x_{\tt bj},b,\mu_h,\zeta_f\right)  & = f_{q/p}\left(x_{\tt bj},b,\mu_b,\zeta_{i}\right) \, U_{\rm NP}(b,Q_0,Q) \nn \\
&\times \exp\left[\int_{\mu_b}^{\mu_h} \frac{\d\mu}{\mu}\gamma^f_\mu\left(\mu,\zeta_f\right)\right]
\left(\frac{\zeta_f}{\zeta_{i}}\right)^{\frac{1}{2}\gamma^f_\zeta(b,\,\mu_b)}\,,
\end{align}
with 
\begin{align}
    f_{q/p}\left(x_{\tt bj},b,\mu_b,\zeta_{i}\right) =\sum_i \int_{x_{\tt bj}}^1 \frac{\d {z }}{z} I_{q / i}(z, b, \mu_b, \zeta_{i} ) f_{i/p}(x_{\tt bj}/z, \mu_b)\,.
\end{align}
Here $I_{q / i} \equiv \mathcal{I}_{q / i}\sqrt{\mathcal S_\perp}$. Besides, we use CT18NNLO \cite{Hou:2019qau} with $\alpha_s(m_Z)=0.118$ for the collinear PDF. 

In figure \ref{fig:res}, we present the resummation results at NNLL and N$^3$LL accuracy for the EIC (left panels) and HERA (right panels). We also illustrate the perturbative uncertainties arising from scale variations. Specifically, the hard scale $\mu_h$ (upper panels) and the soft scale $\mu_b$ (lower panels) are varied up and down by a factor of two around the default values given in Eq.~\eqref{eq:resscales}. The uncertainty bands are narrower at N$^3$LL (red) compared to NNLL (blue), indicating reduced perturbative uncertainties at higher logarithmic accuracy. These bands overlap almost entirely, demonstrating the perturbative convergence of the resummation formula.

\begin{figure}[t]
    \centering
    \includegraphics[scale=0.43]{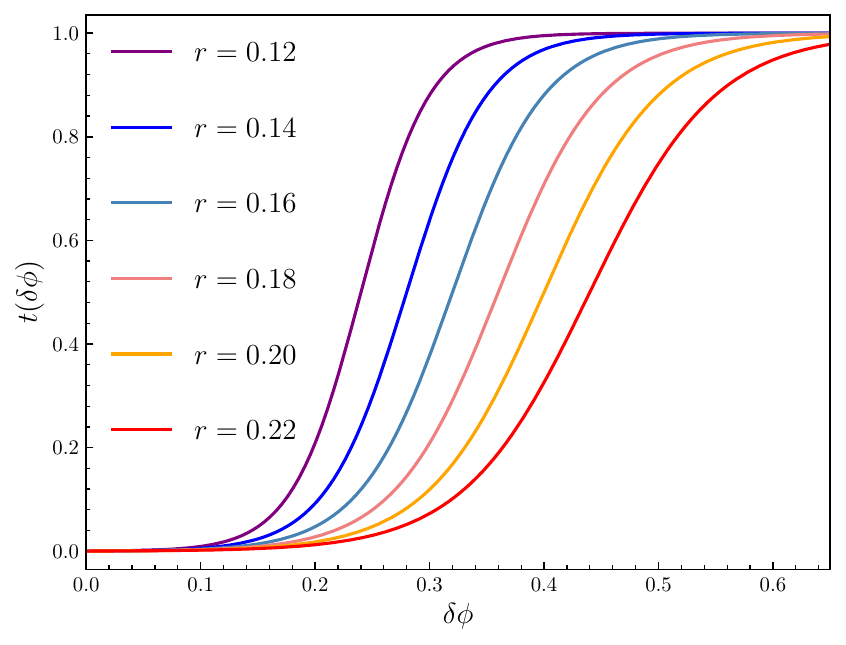}~
    \includegraphics[scale=0.43]{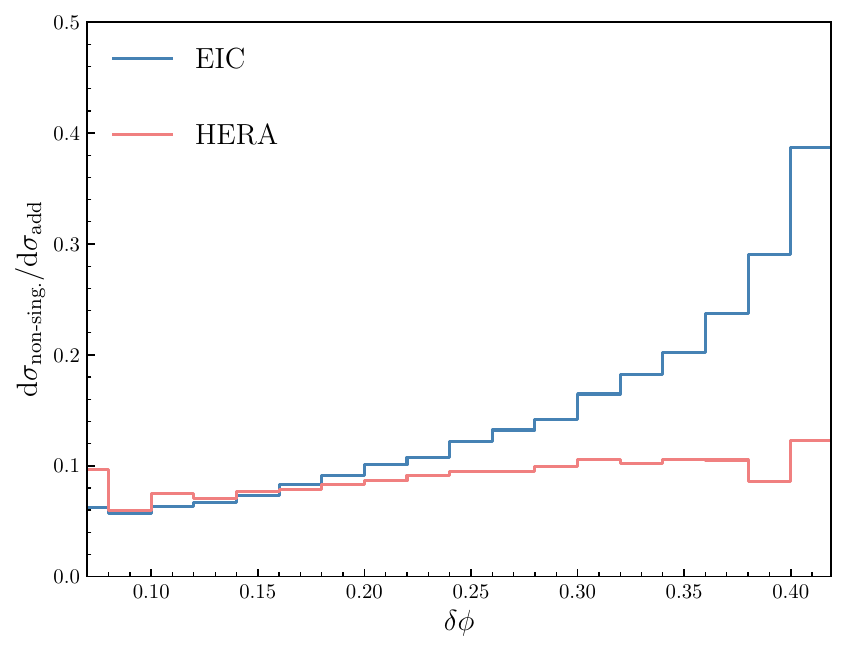} 
    \includegraphics[scale=0.43]{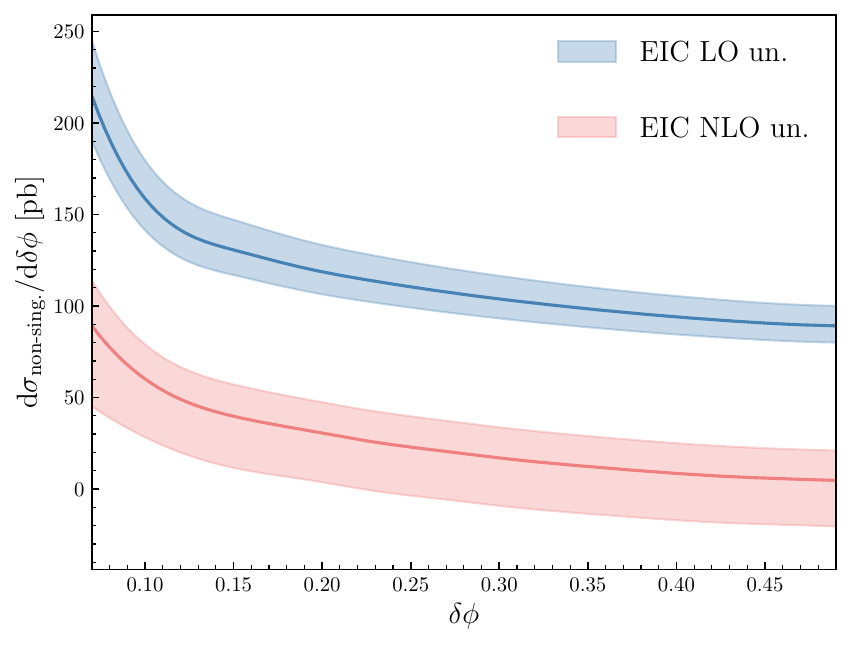}~
    \includegraphics[scale=0.43]{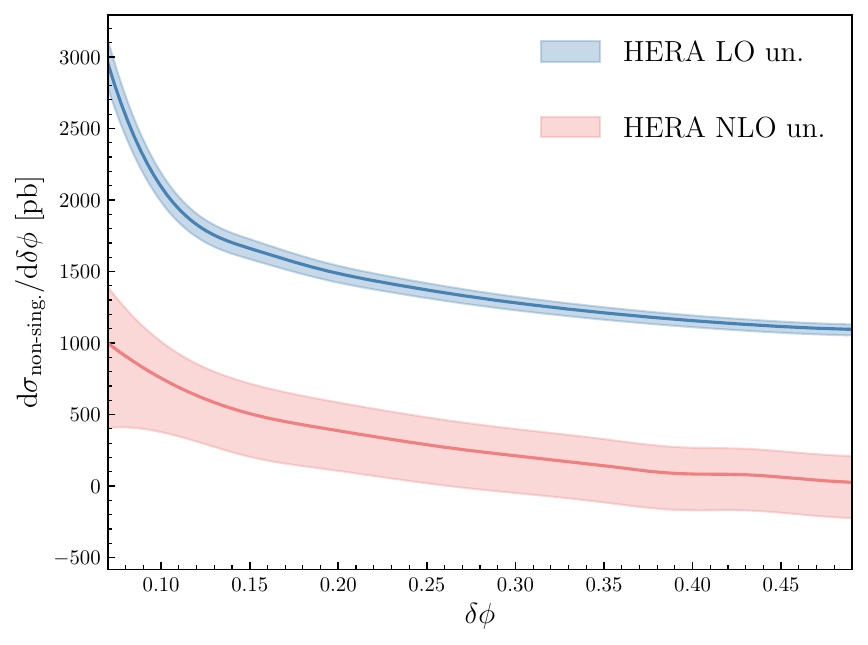}
    \caption{ Upper Left: the transition function $t(\delta\phi)$ with $r=0.12$, $0.14$, $0.16$, $0.18$, $0.20$ and $0.22$. Upper right: the ratio between non-singular and the matched cross section in the additive matching scheme for the EIC (blue) and HERA (red). Lower: the non-singular cross section at LO (blue) and NLO (red) for the EIC (left) and HERA (right). The hard scale  $\mu_h$ is varied from $Q/2$ to $2Q$.}
    \label{fig:prof}
\end{figure}

In the back-to-back limit as $\delta\phi\to 0$, the resummation formula \eqref{eq:res} mitigates the divergent behavior observed in fixed-order results by resumming large logarithms. However, for larger values of $\delta\phi$, the resummation formula receives significant corrections proportional to powers of $\delta\phi$. In this region, it becomes necessary to switch off the resummation and instead employ fixed-order calculations to accurately incorporate these matching corrections.

In the additive matching scheme, the matched cross section for the resummation results and the fixed-order results is defined as
\begin{align}
    \mathrm{d} \sigma_{\text {add }}(\mathrm{N^3LL}+\mathcal{O}(\alpha_s^2))\equiv\mathrm{d} \sigma(\mathrm{N^3LL})+\underbrace{\mathrm{d} \sigma(\mathrm{NLO})-\mathrm{d} \sigma(\mathrm{NLO} \text { singular})}_{\mathrm{d} \sigma(\mathrm{NLO} \text { non-singular})}\,,
\end{align}
where the NLO non-singular distribution is the difference between NLO and NLO singular results. In this work, we use the {\tt NLOJET++} generator to calculate the fixed-order results. In the large $\delta\phi$ region, to avoid numerical instability of the resummation formula, we apply a transition function $t(\delta \phi)$ to obtain the complete matched cross section at N$^3$LL + $\mathcal{O}(\alpha_s^2)$ accuracy, defined as 
\begin{align}\label{eq:prof_matching}
    \mathrm{d} \sigma(\mathrm{N^3LL}+\mathcal{O}(\alpha_s^2))=[1-t(\delta \phi)] \mathrm{d} \sigma_{\text {add }}(\mathrm{N^3LL}+\mathcal{O}(\alpha_s^2))+t(\delta \phi) \mathrm{d} \sigma(\mathrm{NLO})\,,
\end{align}
where the transition function is given by
\begin{align}\label{eq:transition_function}
    t(\delta \phi) = \frac{1}{2}-\frac{1}{2}\tanh\left(4-\frac{2\delta\phi}{r}\right)\,.
\end{align}
Here the parameter $r$ determines the transition point, approximately at $r \sim \delta\phi/2$. In the upper left panel of figure \ref{fig:prof}, we present the transition function $t(\delta\phi)$ with $r=0.12$, $0.14$, $0.16$, $0.18$, $0.20$ and $0.22$. In the upper right panel, we show the ratio between the non-singular and the matched cross section in the additive scheme for the EIC (blue) and HERA (red). We observe significant matching corrections in the EIC, exceeding $20\%$ for $\delta \phi \geq 0.3$, while at HERA the matching corrections are milder, around $10\%$ as $\delta \phi < 0.4$.  The lower panels of figure \ref{fig:prof} show the LO (blue) and NLO (red) non-singular contributions along with their associated scale uncertainties. A significant cancellation between the LO and $\delta$NLO non-singular terms is observed, leading to a substantial reduction in the NLO non-singular contributions relative to the LO ones. As a result, the relative scale uncertainties for the NLO non-singular distributions are considerably larger. Understanding the behavior of higher-order corrections to the non-singular contributions, including $\mathcal{O}(\alpha_s^3)$ fixed-order corrections and the resummation of next-to-leading power corrections, remains an important area for further study. A detailed exploration of these effects will be addressed in future work.

\begin{figure}[t]
    \centering
    \includegraphics[scale=0.48]{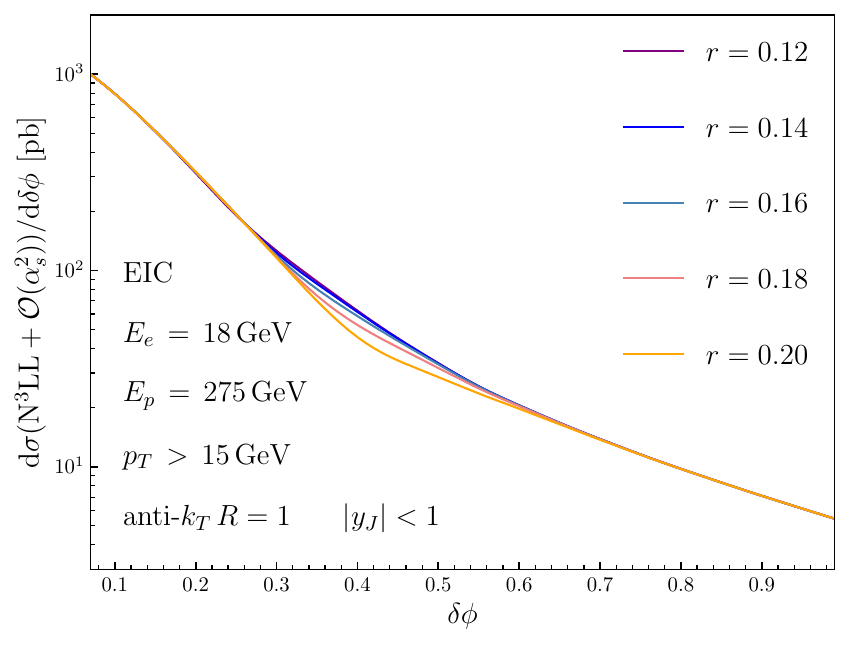}~
    \includegraphics[scale=0.48]{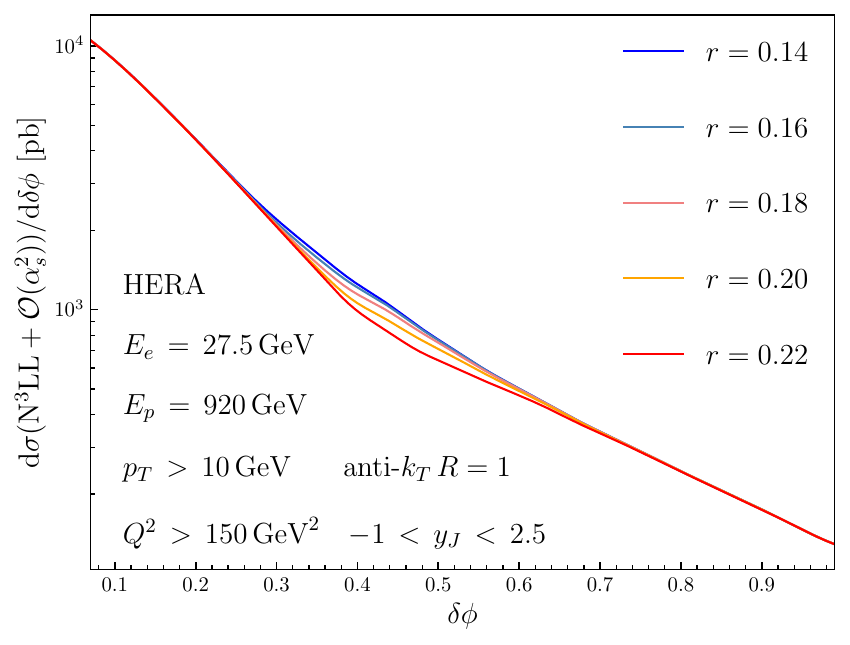} 
    \caption{ 
    Varying the transition point $r$ for the N$^3$LL + $\mathcal{O}(\alpha_s^2)$ results of the $\delta \phi$ distribution at the EIC (left) and HERA (right). The values of $r$ are indicated in the figure.
    }
    \label{fig:N3LLNLO_r}
\end{figure}
\begin{figure}[t]
    \centering
    \includegraphics[scale=0.48]{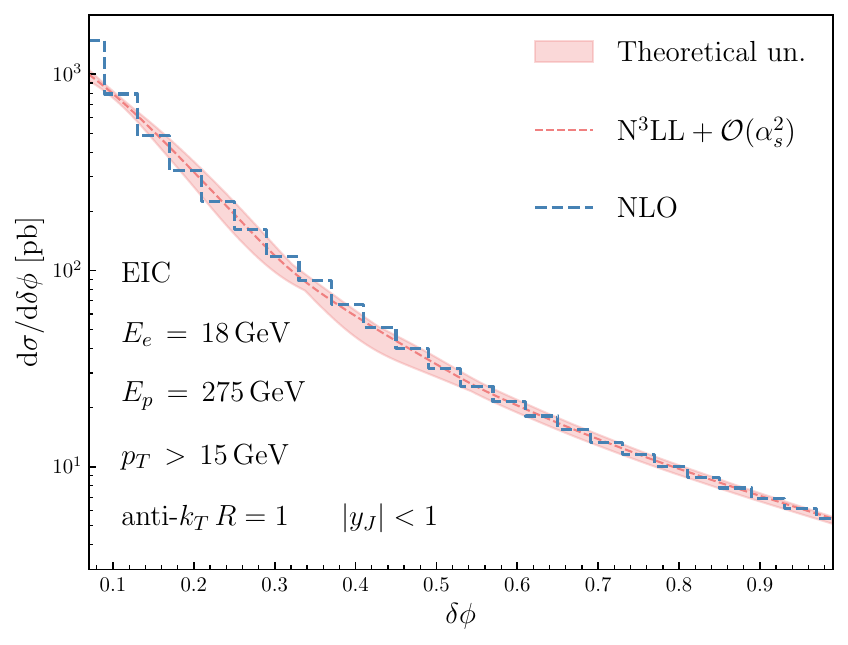}~
    \includegraphics[scale=0.48]{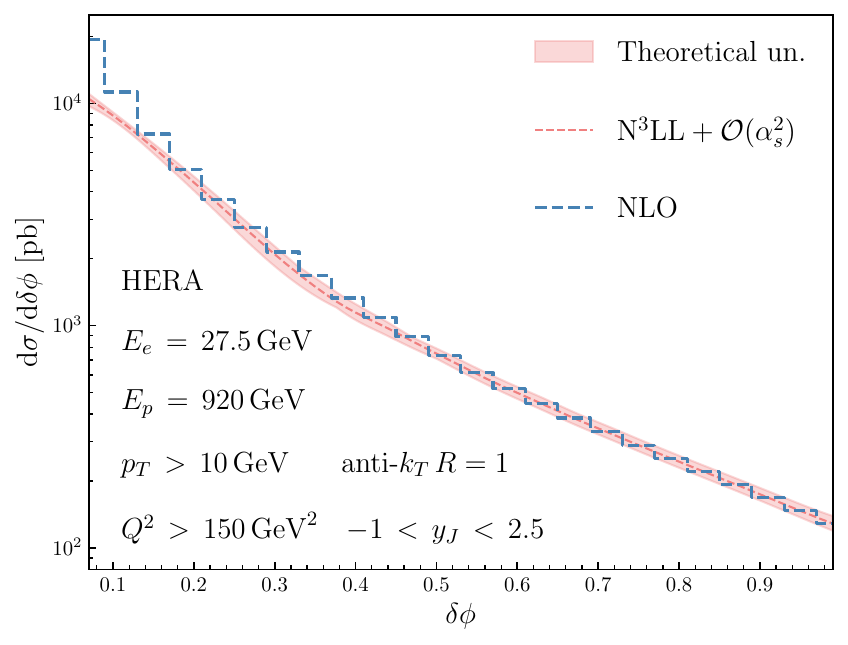} 
    \caption{N$^3$LL + $\mathcal{O}(\alpha_s^2)$ predictions for the $\delta\phi$ distribution at the EIC (left) and HERA (right). The shaded bands indicate theoretical uncertainties, which include scale variations and uncertainties from the transition function. Scale uncertainties are evaluated by varying the hard and soft scales within $0.5\,Q < \mu_h < 2\,Q$ and $0.5\,b_0/b_* < \mu_b < 2\,b_0/b_*$, and are combined in quadrature. Additionally, we vary the transition parameter $r$ in Eq.~\eqref{eq:transition_function} over the ranges $0.12 < r < 0.20$ for the EIC and $0.14 < r < 0.22$ for HERA. Our results are compared to the NLO predictions (blue dashed).
    }
    \label{fig:N3LLNLO}
\end{figure}

In figure \ref{fig:N3LLNLO_r}, we present the dependence of the N$^3$LL + $\mathcal{O}(\alpha_s^2)$ result on the choice of the transition point $r$ for the EIC (left) and HERA (right). We select $r = 0.16$ for the EIC and $r = 0.18$ for HERA as the default transition points in the transition function. In figure \ref{fig:N3LLNLO}, we present predictions at N$^3$LL + $\mathcal{O}(\alpha_s^2)$ accuracy for the EIC (left) and HERA (right). The theoretical uncertainties are represented by red bands, obtained by varying the scales $\mu_b$ and $\mu_h$ within the ranges $0.5\, b_0/b_*<\mu_b<2\,b_0/b_*$ and $0.5\,Q<\mu_h<2\,Q$, respectively, and varying the transition points within the ranges $0.12<r<0.20$ for the EIC and $0.14<r<0.22$ for HERA. The scale uncertainties are combined in quadrature. In the small $\delta\phi$ region, the dominant uncertainties arise from both the $\mu_h$ variation and the $\mu_b$ variation, with both contributions being comparable. In the transition region $\delta\phi$ from $0.3$ to $0.6$, the uncertainties are primarily dominated by $r$ variation. In the large $\delta\phi$ region, the uncertainties are primarily dominated by $\mu_h$ variation. Additionally, in figure \ref{fig:N3LLNLO} our predictions are also compared with the NLO results (blue dashed), which exhibit divergences as $\delta\phi \to 0$. Incorporating the matching corrections given in Eq.~\eqref{eq:prof_matching} ensures that our predictions converge smoothly to the NLO results at larger values of $\delta\phi$, where resummation becomes unnecessary. These numerical predictions provide a pathway for precision studies of QCD and the inner structure of nucleons in DIS, and are particularly relevant for analyses involving HERA data and forthcoming EIC data.

\section{Conclusion} \label{sec:conclusion}

In this paper, we have presented a detailed analysis of the azimuthal angular distribution between the lepton and the leading jet in DIS at N$^{3}$LL + $\mathcal{O}(\alpha_s^2)$ accuracy. Our study utilized the TMD factorization formalism within the framework of SCET. By employing the anti-$k_T$ clustering algorithm ($R=1$) and the winner-take-all recombination scheme, we properly defined the final-state jets in the lab frame. Additionally, we validated the extraction of the two-loop constant terms in the jet function through numerical methods and cross-checked our results with fixed-order predictions for the back-to-back dijet process in $e^+e^-$ collisions. Furthermore, the comparison between our resummation results and fixed-order calculations emphasizes the necessity of incorporating matching corrections to achieve accurate predictions across different kinematic regions. The successful application of the transition function to smoothly interpolate between the resummation and fixed-order regimes demonstrates the robustness of our approach.

Our findings highlight the importance of precision QCD calculations in reducing theoretical uncertainties and enhancing the reliability of predictions for experimental observables. The methodologies and results presented herein are expected to significantly aid in the analysis of current and future DIS data, thereby advancing our understanding of the inner structure of nucleons. Looking forward, ongoing and future research efforts aimed at incorporating resummation of power corrections to jet observables will further improve the precision and applicability of TMD factorization and resummation techniques in DIS. Our study provides a foundational step towards these goals and emphasizes the significance of high-order resummation in accurately characterizing hadronic structure.

\acknowledgments
The authors thank Wouter Waalewijn for the helpful discussion. S.F., M.S.G. and D.Y.S. are supported by the National Natural Science Foundation of China under Grant No.~12275052, No.~12147101 and No.~12347147. H.T.L. is supported by the National Science Foundation of China under grant Nos. 12275156 and 12321005. M.S.G. is also supported by China Postdoctoral Science Foundation under Grant No. 2023M740720.

\appendix 

\section{Resummation formula in $e^+e^-$ and the two-loop jet function}\label{app:epem}

This appendix briefly reviews the factorization and resummation formula for the transverse momentum decorrelation in the $e^+e^- \to $ dijet process \cite{Gutierrez-Reyes:2019vbx}. We follow this by outlining the fixed-order expansion of the resummed expressions and present the methods for the numerical determination of two-loop constants in the jet function, utilizing the {\tt NLOJET++} generator.

In the coordinate $b$-space, the factorization theorem reads
\begin{align}\label{eq:epem_fac}
    \frac{\d\sigma}{\d^2 \bm q_{T}}& =
     \bar\sigma_0 \, H(Q,\mu) \!\! \int \!\! \frac{\d^2\bm{b}_T}{(2\pi)^2} e^{i \bm{q}_{T}\cdot \bm{b}_T}  {\cal J}_q( b_T,\mu,\zeta/\nu^2) {\cal J}_{\bar q}( b_T,\mu,\zeta/\nu^2) S( b_T,\mu,\nu)\,, 
\end{align}
where the Born cross section is given by
\begin{align}
    \bar\sigma_0 = \frac{4\pi\alpha_{\rm em}^2 }{Q^2} \sum_{f} e_f^2\,.
\end{align}
In the standard CSS treatment \cite{Collins:2011zzd}, the genuine jet function is defined as
\begin{align}
    J_{q}( b_T,\mu,\zeta) \equiv {\cal J}_{q}(b_T,\mu,\zeta/\nu^2) \sqrt{S( b_T,\mu,\nu)}\,,  
\end{align}
where the rapidity scale $\nu$ dependence cancels once two contributions are combined. 

At the renormalization scale $\mu$, after solving the CSS evolution equation for the $\zeta$ dependence in the jet function, we obtain 
\begin{align}
    J_q( b_T,\mu,\zeta_f) = J_q( b_T,\mu,\zeta_i) \left(\frac{\zeta_f}{\zeta_{i}}\right)^{\frac{1}{2} \gamma_\zeta^J\left(b_T, \mu\right)}\,.
\end{align}
In addition to the CSS evolution equation, we also consider the standard RG evolution between the hard function and TMD jet functions, with the corresponding RG equations:
\begin{align}
     \frac{\d}{\d\,\ln \mu} H(Q,\mu) &= \left[2\,C_F\gamma _{\text{cusp}}( \alpha _{s})\,\mathrm{ln}\frac{Q^{2}}{\mu ^{2}} +2\gamma _{V}( \alpha _{s})\right] H(Q,\mu)\,, \\
        \frac{\mathrm{d}}{\mathrm{d}\,\ln{\mu}}J_q(b_T,\mu,\zeta_f)&=\left[C_F\gamma_{\rm cusp}(\alpha_s)\ln{\frac{\mu^2}{\zeta_f}}+\gamma_J(\alpha_s)\right]J_q(b_T,\mu,\zeta_f)\,.  
\end{align}
The above RG equations hold to all orders in perturbative QCD as a consequence of the factorization theorem in Eq.~\eqref{eq:epem_fac}. Therefore, the cusp anomalous dimension terms imply that $\zeta_f=Q^2$, where we make the Collins-Soper scales in the two TMD jet functions equal. Besides, the RG consistency implies that 
\begin{align}
    \gamma_{V}(\alpha_s) + \gamma_{J}(\alpha_s)  = 0\,. 
\end{align}
The perturbative coefficients of all anomalous dimensions needed are summarized in the appendix \ref{app:adim}.

After evolving the jet function from $\{\mu_b,\zeta_i\}$ to $\{\mu_h,\zeta_f\}$, we obtain
\begin{align}
    J_{q}(b_T,\mu_h,\zeta_f) = e^{-S_{\rm pert}(\mu_{b}, \, \mu_h)} \left(\frac{\zeta_f}{\zeta_{i}}\right)^{\frac{1}{2} \gamma_\zeta^J\left(b_T, \, \mu_b\right)} J_{q}(b_T,\mu_{b},\zeta_i)\,, 
\end{align}
with the perturbative Sudakov factor given by
\begin{align}
    S_{\rm pert}(\mu_{b}, \, \mu_h) = -\int_{\mu_b}^{\mu_h} \frac{\d  \mu}{ \mu} \left[C_F\gamma_{\rm cusp}(\alpha_s)\ln{\frac{\mu^2}{\zeta_f}}+\gamma_J(\alpha_s)\right]\,,
\end{align}
where the renormalization and the Collins-Soper scales are chosen as
\begin{align}
    \mu_h=Q,\quad\quad\mu_b=b_0/b_T,\quad\quad\zeta_i = b_0^2/b_T^2,\quad\quad\zeta_f=Q^2\,.
\end{align}
The all-order resummed cross section is written as
\begin{align}\label{eq:epem_res}
    \frac{\d\sigma}{ \d q_{T}}=
     {\bar \sigma}_0 \, H(Q,\mu_h) \, q_T \! \int_0^\infty  b_T \, \d b_T  J_0( q_{T} b_T)  J_{q}(b_T,\mu_h,\zeta_f) J_{\bar q}(b_T,\mu_h,\zeta_f)\,,  
\end{align}
where $J_0$ is the Bessel function of the first kind.

It is instructive to perform the fixed-order expansion of the resummed result in Eq. \eqref{eq:epem_res}. To this end, we first define the integrated cross section in the form
\begin{align}\label{eq:epem_intres}
    \sigma_L(Q_T) \equiv \int_0^{Q_T} \d q_T \, \frac{\d \sigma}{\d q_T}\,. 
\end{align}
Here the perturbative expansion of the integrated cross section $\sigma_L$ can be expressed as
\begin{align}
    \frac{\sigma_L(Q_T)}{\bar\sigma_0} = 1 + \frac{\alpha_s(Q)}{2\pi} A(Q_T) + \left[\frac{\alpha_s(Q)}{2\pi}\right]^2 B(Q_T) + \mathcal{O}(\alpha_s^3)\,, 
\end{align}
where the two-loop coefficient $B$ is given by
\begin{align}
B &=  \, C_F^2 \Bigg[\frac{\bar L^4}{2} + 3 \bar L^3 + \bar L^2 \left(\frac{11}{2} - \frac{\pi^2}{3} + 6 \, \ln 2\right) + \bar L \left(\frac{9}{4} + 18 \, \ln 2 - 4 \, \zeta_3\right) - \frac{189}{16} + 5 \,\pi^2  \notag \\
& - \frac{173 \, \pi^4}{720} + 27 \,\ln 2 - \frac{9}{2} \pi^2 \,\ln 2 + 9 \, \ln^2 2 - 3 \, \zeta_3\Bigg] +
C_F\,C_A \Bigg[\frac{11 \bar  L^3}{9} + \bar L^2 \left(-\frac{35}{36} + \frac{\pi^2}{6}\right)   \notag\\
&+ \bar L \left(-\frac{57}{4} + \frac{11 \pi^2}{18} + 11 \, \ln 2 + 6 \, \zeta_3\right) - \frac{51157}{1296}  + \frac{1061 \pi^2}{216} - \frac{2 \pi^4}{45} + \frac{401 \zeta_3}{18}\Bigg] \notag\\
&+
C_F\,T_F\,n_f \Bigg[-\frac{4 \bar L^3}{9} + \frac{\bar L^2}{9} + \bar L \left(5 - \frac{2 \pi^2}{9} - 4 \, \ln 2\right) + \frac{4085}{324} - \frac{91 \pi^2}{54} - \frac{14 \zeta_3}{9}\Bigg] \notag \\
&+ \frac{j_2}{2}\,,
\end{align}
where we define $\bar L\equiv \ln \, Q_T^2/Q^2$ and $j_2$ denotes the two-loop constant term in the jet function, which depends on the clustering algorithm and recombination scheme in the jet definition.

\begin{figure}[t]
    \centering
    \includegraphics[scale=0.65]{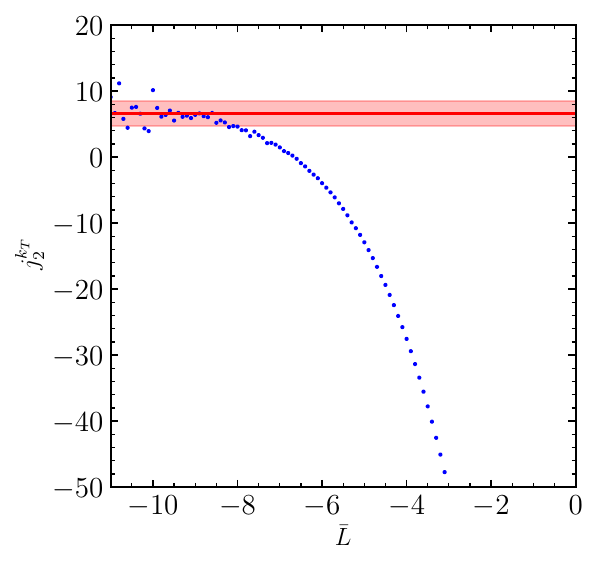}~
    \includegraphics[scale=0.65]{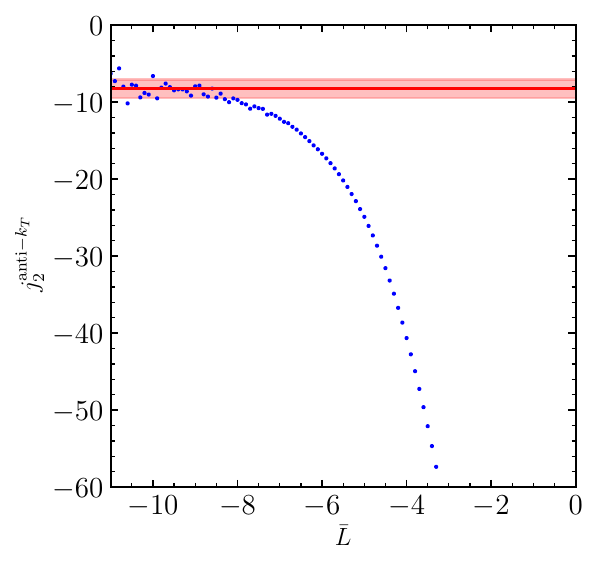}
    \caption{ Determination of the two-loop constant $j_2$ in the $k_T$ (left) and anti-$k_T$ (right) algorithms.
    }
    \label{fig:j2}
\end{figure}

Using the above results, we can determine the two-loop constant term $j_2$. To do so, we compute
\begin{align}
   R(Q_T) \equiv  1 - \int_{Q_T}^{Q_{T,\,{\rm max}}} \d q_T \, \frac{1}{\sigma} \frac{\d \sigma}{\d q_T}\,,
\end{align}
where the ingredient on the right-hand side can be calculated using {\tt NLOJET++} numerically. Here, $\sigma$ denotes the total hadronic cross section, and in the $\overline{\mathrm{MS}}$ scheme, its perturbative expansion reads
\begin{align}
    \frac{\sigma}{\bar\sigma_0}=1+\frac{ \alpha_s(Q)}{ \pi} + \left[\frac{\alpha_s(Q)}{\pi}\right]^2 C_2 +  \mathcal{O}(\alpha_s^3)\,, 
\end{align}
with the two-loop coefficient given by
\begin{align}
    C_2 = \frac{365}{24} - 11\zeta_3 + \left( \frac{2}{3}\zeta_3 - \frac{11}{12} \right)n_f\,.
\end{align}
We then consider the difference between the full result $R(Q_T)$ and its logarithmic part $R_L(Q_T)\equiv\sigma_L(Q_T)/\sigma$ obtained from the Eq.~\eqref{eq:epem_intres}. Since all two-loop ingredients except the constant $j_2$ are known, the constant follows immediately from the requirement that the difference $R-R_L$ must vanish in the limit $Q_T\ll Q$. Figure \ref{fig:j2} shows the difference as a function of the logarithm as well as the value of $j_2$ we extract from it. Based on the above methods, we obtain
\begin{align}
    j_2^{k_T} = 6.60 \pm 1.88\,,\qquad j_2^{\mathrm{anti}{\text -}k_T} = -8.25 \pm 1.18\,.
\end{align}

In table \ref{tab:j_2}, we present a comparison of our results for the $j_2$ constants with those obtained from GSWZ \cite{Gutierrez-Reyes:2019vbx}, based on numerical fitting using {\tt Event2}, and BBDW \cite{Brune:2022cgr, scet2023}, obtained from direct numerical calculations. For the $k_T$ algorithm, our result is lower than both GSWZ and BBDW. Conversely, for the anti-$k_T$ algorithm, our result aligns more closely with BBDW than with GSWZ.

\begin{table}[t]
\centering 
\begin{tabular}{|c|c|c|c|}
\hline
$j_2$   & This work &  GSWZ\    &  BBDW\          \\ \hline
$k_T$   & $6.60 \pm 1.88$ & $10.1\pm1.2$      & $13.28\pm1.47$          \\ \hline
$\textrm{anti-}k_T$   & $-8.25\pm 1.18$  &$-6.5\pm0.7$      & $-9.91\pm0.88$             \\ \hline
\end{tabular}
\caption{Comparison of numerical results for the two-loop constant $ j_2 $  obtained in this work with those from previous studies (GSWZ \cite{Gutierrez-Reyes:2019vbx} and BBDW \cite{Brune:2022cgr, scet2023}).}
\label{tab:j_2}
\end{table}

\section{Anomalous dimensions and resummation ingredients}\label{app:adim}

All the anomalous dimensions in Eq.~\eqref{eq:AD_jet} have the following perturbative expansion in the strong coupling constant
\begin{align}
    \gamma_i(\alpha_s) &= \sum_{n=0}^\infty\gamma^i_n\left(\frac{\alpha_s}{4\pi}\right)^{n+1}\,, 
    \quad{\rm with} \quad i={\rm cusp},f,J,r,
\end{align}
and at $\rm N^3LL$ the coefficients are given by~\cite{Korchemsky:1987wg, Moch:2004pa, Moch:2005id, Moch:2005tm, Becher:2006mr, Henn:2019swt}
\begin{align}
  \gamma_{0}^{\rm cusp} &=  \, 4 \,,
  \\
  \gamma_{1}^{\rm cusp} &=  \, 4  \left[ C_A\left(\frac{67}{9}-\frac{\pi^2}{3}\right)-\frac{20  }{9}T_F n_f\right] \nn \,,
  \\
  \gamma_{2}^{\rm cusp} &=  \, 4\Bigg[C_A^2  \left(\frac{245}{6}-\frac{134\pi^2}{27}+\frac{11\pi^4}{45}+\frac{22
      \zeta_3}{3}\right) 
      +C_A T_F n_f \left(-\frac{418}{27}+\frac{40\pi^2}{27}-\frac{56
    \zeta_3}{3}\right) \nn \\
    & + C_F T_F n_f \left(-\frac{55}{3}+16 \zeta_3\right)
    -\frac{16}{27}T_F^2 n_f^2 \Bigg]\,,\nn\\
\gamma_{3}^{\rm cusp} &=  256\Bigg[C_A^3\left(\frac{1309\zeta_3}{432}-\frac{11\pi^2\zeta_3}{144}-\frac{\zeta_3^2}{16}-\frac{451\zeta_5}{288}+\frac{42139}{10368}-\frac{5525\pi^2}{7776}+\frac{451\pi^4}{5760}-\frac{313\pi^6}{90720}\right)\nn\\
&+C_A^2 T_F n_f\left(-\frac{361\zeta_3}{54}+\frac{7\pi^2\zeta_3}{36}+\frac{131\zeta_5}{72}-\frac{24137}{10368}+\frac{635\pi^2}{1944}-\frac{11\pi^4}{2160}\right)\nn\\
&+C_F C_A T_F n_f\left(\frac{29\zeta_3}{9}-\frac{\pi^2\zeta_3}{6}+\frac{5\zeta_5}{4}-\frac{17033}{5184}+\frac{55\pi^2}{288}-\frac{11\pi^4}{720}\right)\nn\\
&+C_F^2 T_F n_f\left(\frac{37\zeta_3}{24}-\frac{5\zeta_5}{2}+\frac{143}{288}\right)
+C_A T_F^2 n_f^2\left(\frac{35\zeta_3}{27}-\frac{19\pi^2}{972}-\frac{7\pi^4}{1080}+\frac{923}{5184}\right)\nn\\
&+C_F T_F^2 n_f^2\left(-\frac{10\zeta_3}{9}+\frac{\pi^4}{180}+\frac{299}{648}\right)
+T_F^3 n_f^3\left(\frac{2\zeta_3}{27}-\frac{1}{81}\right)\nn\\
&+\frac{d_F^{abcd}d_A^{abcd}}{C_F N_c}\left(\frac{\zeta_3}{6}-\frac{3\zeta_3^2}{2}+\frac{55\zeta_5}{12}-\frac{\pi^2}{12}-\frac{31\pi^6}{7560}\right)
+n_f\frac{d_F^{abcd}d_F^{abcd}}{C_F N_c}\left(-\frac{\zeta_3}{3}-\frac{5\zeta_5}{3}+\frac{\pi^2}{6}\right)
\Bigg]\,,\nn
\end{align}
with
\begin{align}
\frac{d_F^{abcd}d_A^{abcd}}{C_F N_c}=\frac{N_c(N_c^2+6)}{24}\,,\qquad
\frac{d_F^{abcd}d_F^{abcd}}{C_F N_c}=\frac{N_c^4-6N_c^2+18}{48N_c^2}\,,
\end{align}
and
\begin{align}
  \gamma^{f/J}_0 & =   6\, C_F \,, \\
  \gamma^{f/J}_1 & =   C_F^2\left(3-4 \pi^2+48 \zeta_3\right)+C_F C_A\left(\frac{961}{27}+\frac{11 \pi^2}{3}-52 \zeta_3\right)+C_F T_F n_f\left(-\frac{260}{27}-\frac{4 \pi^2}{3}\right) \,, \nn \\
  \gamma^{f/J}_2 & =C_F^3 \left( 29 + 6 \pi^2 + \frac{16 \pi^4}{5} + 136 \zeta_3 - \frac{32 \pi^2 \zeta_3}{3} - 480 \zeta_5 \right)\nn\\
  &+ C_F^2 C_A \left( \frac{151}{2} - \frac{410 \pi^2}{9} - \frac{494 \pi^4}{135} + \frac{1688 \zeta_3}{3} + \frac{16 \pi^2 \zeta_3}{3} + 240 \zeta_5 \right)\nn\\
  &+ C_F C_A^2 \left( \frac{139345}{1458} + \frac{7163 \pi^2}{243} + \frac{83 \pi^4}{45} - \frac{7052 \zeta_3}{9} + \frac{88\pi^2 \zeta_3}{9} + 272 \zeta_5 \right)\nn\\
&+ C_F^2 T_F n_f \left( -\frac{5906}{27} + \frac{52 \pi^2}{9} + \frac{56 \pi^4}{27} - \frac{1024 \zeta_3}{9} \right)\nn\\
&+ C_F C_A T_F n_f \left( \frac{34636}{729} - \frac{5188 \pi^2}{243} - \frac{44 \pi^4}{45} + \frac{3856 \zeta_3}{27} \right)\nn\\
&+ C_F T_F^2 n_f^2 \left( -\frac{19336}{729} + \frac{80 \pi^2}{27} + \frac{64 \zeta_3}{27} \right)
\,.\nn
\end{align}
Besides, the rapidity anomalous dimensions in Eq.~\eqref{eq:AD_jet} are given by \cite{Li:2016ctv}
\begin{align}
\gamma_0^r &= 0 \,, \\
\gamma_1^r &=  C_F C_A \left(-\frac{808}{27} + 28 \zeta_3\right) + \frac{224}{27}C_F T_F n_f  \,,\nn\\
\gamma_2^r &= 
  C_F C_A^2\left( -\frac{297029}{729} + \frac{3196 \pi^2}{243} + \frac{77 \pi^4}{135} + \frac{12328 \zeta_3}{27} - \frac{88\pi^2 \zeta_3}{9}  - 192 \zeta_5 \right)\nn\\
&+C_F^2 T_F n_f  \left( \frac{3422}{27} - \frac{16 \pi^4}{45} - \frac{608 \zeta_3}{9} \right)
+C_F T_F^2 n_f^2  \left( -\frac{7424}{729} - \frac{128 \zeta_3}{9} \right)\nn\\
&+C_F C_A T_F n_f  \left( \frac{125252}{729} - \frac{824 \pi^2}{243} + \frac{4\pi^4}{27} - \frac{1808 \zeta_3}{27} \right) \,.\nn
\end{align}
The hard function is given by 
\begin{align}
 H(Q,\mu)=|C(Q,\mu)|^2 \,,  
\end{align}
where $C(Q,\mu)$ is the Wilson coefficient, which up to NNLO is expressed as \cite{Matsuura:1987wt,Matsuura:1988sm,Gehrmann:2005pd,Moch:2005id}
\begin{align}
C(Q,\mu)=1+\frac{\alpha_s C_F}{4\pi}\left(-L_\mu^2+3L_\mu-8+\frac{\pi^2}{6}\right)+\left(\frac{\alpha_s}{4\pi}\right)^2 C_F(C_F H_F+C_A H_A+ T_F n_f H_f)\,,
\end{align}
with
\begin{align}
H_F&=\frac{L_\mu^4}{2} - 3L_\mu^3 + L_\mu^2\left( \frac{25}{2}-\frac{\pi^2}{6} \right) 
+ L_\mu\left(-\frac{45}{2}-\frac{3\pi^2}{2}+24\zeta_3 \right)\nn\\
&+\frac{255}{8} + \frac{7 \pi^2}{2} - \frac{83 \pi^4}{360} - 30 \zeta_3,\nn \\
H_A&=\frac{11}{9}L_\mu^3+L_\mu^2\left(-\frac{233}{18}+\frac{\pi^2}{3}\right)
+L_\mu\left(\frac{2545}{54}+\frac{11\pi^2}{9}-26\zeta_3\right)\nn\\
&-\frac{51157}{648} - \frac{337 \pi^2}{108} + \frac{11\pi^4}{45} + \frac{313}{9} \zeta_3
\,, \nn\\
H_f&=-\frac{4}{9}L_\mu^3+\frac{38}{9}L_\mu^2+
L_\mu\left(-\frac{418}{27}-\frac{4\pi^2}{9}\right)
+\frac{4085}{162} + \frac{23 \pi^2}{27} + \frac{4}{9}\zeta_3\,,  
\end{align}
and $L_\mu=\ln\left( Q^2/\mu^2\right)$ for SIDIS or $L_\mu=\ln\left( Q^2/\mu^2\right)-i\pi$ for $e^+e^-$. The perturbative expansion of the matching coefficient in Eq.~\eqref{eq:OPE} can be expressed as:
\begin{align}
\mathcal{I}_{q / i}(z, b, \mu, \nu ) &= \sum_{n=0}^\infty\mathcal{I}_{q/i}^{(n)}\left(z,b,L_Q\right)\left(\frac{\alpha_s}{4\pi}\right)^{n}\,, 
\end{align}
with two logarithms
\begin{align}
    L_b=\ln\frac{ \mu^2 b^2}{b_0^2}\,, 
    \qquad
    L_Q=2\,\ln \frac{x_{\tt bj}(\bar n \cdot P)}{\nu}\,.
\end{align}
Up to $\mathcal{O}(\alpha_s^2)$, we have
\begin{align}
\mathcal{I}_{q/i}^{(0)}\left(z,b,L_Q\right)&=\delta_{qi}\delta(1-z)\,,\\
\mathcal{I}_{q/i}^{(1)}\left(z,b,L_Q\right)&=\delta_{qi}\delta(1-z)
\left(-\frac{\gamma_{0}^{\rm cusp}}{2}C_F L_b L_Q+\frac{\gamma_{0}^{\mathcal{B}}}{2}L_b+\frac{\gamma_{0}^{r}}{2}L_Q\right)
-P_{qi}^{(0)}(z)L_b+I_{qi}^{(1)}(z)\,,\notag\\
\mathcal{I}_{q/i}^{(2)}\left(z,b,L_Q\right)&=\delta_{qi}\delta(1-z)
\Bigg[\frac{1}{8}\left(-\gamma_{0}^{\rm cusp} C_F L_Q
+\gamma_{0}^{\mathcal{B}}\right)\left(-\gamma_{0}^{\rm cusp} C_F L_Q
+\gamma_{0}^{\mathcal{B}}+2\beta_0\right)L_b^2 \nn\\
&+\left(-\frac{\gamma_{1}^{\rm cusp}}{2}C_F L_Q+\frac{\gamma_{1}^{\mathcal{B}}}{2}
+\left(-\gamma_{0}^{\rm cusp} C_F L_Q
+\gamma_{0}^{\mathcal{B}}+2\beta_0\right)\frac{\gamma_{0}^{r}}{4}L_Q\right)L_b\nn\\
&+\frac{(\gamma_{0}^{r})^2}{8}L_Q^2+\frac{\gamma_{1}^{r}}{2}L_Q\Bigg] +\Bigg[\frac{1}{2}\sum_j P_{qj}^{(0)}(z)\otimes P_{ji}^{(0)}(z)
+\frac{P_{qi}^{(0)}(z)}{2}(\gamma_{0}^{\rm cusp} C_F L_Q\nn\\
&-\gamma_{0}^{\mathcal{B}}-\beta_0)\Bigg]L_b^2 
+\Bigg[-P_{qi}^{(1)}(z)-\frac{P_{qi}^{(0)}(z)}{2}\gamma_{0}^{r}L_Q-\sum_j I_{qj}^{(1)}(z)\otimes P_{ji}^{(0)}(z)\nn\\
    &+\left(-\frac{\gamma_{0}^{\rm cusp}}{2} C_F L_Q
+\frac{\gamma_{0}^{\mathcal{B}}}{2}+\beta_0\right)I_{qi}^{(1)}(z)\Bigg]L_b
+\frac{\gamma_{0}^{r}}{2} L_Q I_{qi}^{(1)}(z)
+I_{qi}^{(2)}(z)\,,\nn
\end{align}
with 
\begin{align}
 \gamma_0^{\mathcal{B}}&=  6 C_F \,,   \\
 \gamma_1^{\mathcal{B}}&=  C_F^2\left(3-4 \pi^2+48 \zeta_3\right)+C_F C_A\left(\frac{17}{3}+\frac{44 \pi^2}{9}-24 \zeta_3\right)+C_F T_F n_f\left(-\frac{4}{3}-\frac{16 \pi^2}{9}\right)\,,  \nn    
\end{align}
and $\beta_0=11/3\,C_A-4/3\,T_F n_f$. The one-loop and two-loop splitting functions are given by \cite{Altarelli:1977zs,Furmanski:1980cm,Curci:1980uw}
\begin{align}\label{splitting}
P_{qq}^{(0)}(z)&=2C_F\left[\frac{1+z^2}{(1-z)_+}+\frac{3}{2}\delta(1-z)\right]\,, \\
P_{qg}^{(0)}(z)&=2T_F\left[z^2+(1-z)^2\right]\,,\nn\\
P_{qq'}^{(1)}(z)&=C_F T_F\Bigg[-8(z+1)H_{0,0}+\frac{4}{3}(8z^2+15z+3)H_0+\frac{8(1-z)(28z^2+z+10)}{9z}\Bigg]\,,\nn\\
P_{q\bar q}^{(1)}(z)&=P_{qq'}^{(1)}(z)+(C_A C_F-2C_F^2)\Bigg[
4\frac{1+z^2}{1+z}(2H_{-1,0}-H_{0,0}+\zeta_2)-4(z+1)H_0\nn\\
&-8(1-z)\Bigg]\,,\nn\\
P_{qg}^{(1)}(z)&=C_A T_F\Bigg[-8\left[z^2+(1+z)^2\right]H_{-1,0}
-8\left[z^2+(1-z)^2\right]H_{1,1}-8(2z+1)H_{0,0}\nn\\
&+16z(1-z) H_1+\frac{4}{3}(44z^2+24z+3)H_0-\frac{4(218z^3-225z^2+18z-20)}{9z}-16z\zeta_2\Bigg]\nn\\
&+C_F T_F\Bigg[8\left[z^2+(1-z)^2\right](H_{1,0}+H_{1,1}+H_2-\zeta_2)+4(4z^2-2z+1)H_{0,0}\nn\\
&+2(8z^2-4z+3)H_0-16z(1-z) H_1+2(20z^2-29z+14)\Bigg]\nn\\
P_{qq}^{(1)}(z)&=P_{qq'}^{(1)}(z)
+C_A C_F\Bigg[\left(\frac{268}{9}-8\zeta_2\right)\frac{1}{(1-z)_+}
+4\frac{1+z^2}{(1-z)_+}H_{0,0}+\frac{2(5z^2+17)}{3(1-z)}H_0\nn\\
&+4(z+1)\zeta_2+\left(\frac{44\zeta_2}{3}-12\zeta_3+\frac{17}{6}\right)\delta(1-z)-\frac{2}{9}(187z-53)\Bigg]\nn\\
&+C_F T_F n_f\Bigg[-\frac{80}{9}\frac{1}{(1-z)_+}-\frac{8}{3}\frac{1+z^2}{(1-z)_+}H_{0}+\left(-\frac{16\zeta_2}{3}-\frac{2}{3}\right)\delta(1-z)\nn\\
&+\frac{8}{9}(11z-1)\Bigg]
+C_F^2\Bigg[8\frac{1+z^2}{(1-z)_+}(H_{1,0}+H_2)-4(z+1)H_{0,0}+\frac{4(2z^2-2z-3)}{1-z}H_0\nn\\
&+\left(-12\zeta_2+24\zeta_3+\frac{3}{2}\right)\delta(1-z)-20(1-z)\Bigg]\,,\nn
\end{align}
where $q'$ denotes a light quark with flavor different from $q$, and $H_{a_1,\ldots,a_n}\equiv H(a_1,\ldots,a_n;z)$ is the shorthand notation of the Harmonic PolyLogarithms (HPLs) \cite{Remiddi:1999ew}. We use the packages {\tt HPL} \cite{Maitre:2005uu} and {\tt FastGPL} \cite{Wang:2021imw} to compute these functions numerically. The scale-independent coefficients are given by \cite{Luo:2019hmp}
\begin{align}\label{eq:matching_coefficient}
I_{qq}^{(1)}(z)&=2C_F(1-z)\,,\\
I_{qg}^{(1)}(z)&=4T_Fz(1-z)\,,\nn\\
I_{qq'}^{(2)}(z)&=C_F T_F\Bigg[-\frac{8(1-z)(2z^2-z+2)}{3z}(H_{1,0}+\zeta_2)-\frac{2}{3}(8z^2+3z+3)H_{0,0}\nn\\
&+4(z+1)H_{0,0,0}+\frac{4}{9}(32z^2-30z+21)H_0+\frac{2(1-z)(136z^2-143z+172)}{27z}\Bigg]\,,\nn\\
I_{q\bar q}^{(2)}(z)&=I_{qq'}^{(2)}(z)+(C_A C_F-2C_F^2)\Bigg[
-2\frac{1+z^2}{1+z}(4H_{-2,0}-2H_{2,0}-4H_{-1,-1,0}+2H_{-1,0,0}\nn\\
&-H_{0,0,0}-2H_{-1}\zeta_2+\zeta_3)+4(1-z)H_{1,0}+4(z+1)H_{-1,0}-(11z+3)H_0\nn\\
&+2(3-z)\zeta_2-15(1-z)\Bigg]\,,\nn\\     
I_{qg}^{(2)}(z)&=C_A T_F\Bigg[4\left[z^2+(1+z)^2\right](2H_{-2,0}-2H_{-1,-1,0}+H_{-1,0,0}-H_{-1}\zeta_2)\nn\\
&+4\left[z^2+(1-z)^2\right](H_{1,2}+H_{1,1,0}-H_{1,1,1})
+8z(z+1)H_{-1,0}-8z(1-z)H_{1,1}\nn\\
&-\frac{8(1-z)(11z^2-z+2)}{3z}H_{1,0}-16z H_{2,0}+4(2z+1)H_{0,0,0}+8z\zeta_3+2z(4z-3) H_1\nn\\
&+\frac{4}{9}(68z^2-30z+21)H_0-\frac{2}{3}(44z^2-12z+3)H_{0,0}+\frac{8(11z^3-9z^2+3z-2)}{3z}\zeta_2\nn\\
&-\frac{2(298z^3-387z^2+315z-172)}{27z}\Bigg]+C_F T_F\Bigg[4\left[z^2+(1-z)^2\right](H_{2,1}-H_{1,0,0}\nn\\
&+H_{1,1,1}+7\zeta_3)+(-8z^2+12z+1)H_{0,0}-2(4z^2-2z+1)H_{0,0,0}\nn\\
&+8z(1-z)(H_{1,0}+H_{1,1}+H_2-\zeta_2)+(-8z^2+15z+8)H_0-2z(4z-3)H_1\nn\\
&-72z^2+75z-13\Bigg]\,,\nn\\
I_{qq}^{(2)}(z)&=I_{qq'}^{(2)}(z)
+C_A C_F\Bigg[\left(28\zeta_3-\frac{808}{27}\right)\frac{1}{(1-z)_+}
+2\frac{1+z^2}{(1-z)_+}(-2H_{1,2}-2H_{2,0}-H_{0,0,0}\nn\\
&-2H_{1,1,0})+\frac{(z^2-12z-11)}{3(1-z)}H_{0,0}-4(1-z)H_{1,0}-\frac{2(83z^2-36z+29)}{9(1-z)}H_0-2zH_1\nn\\
&+\frac{2(z^2-13)}{1-z}\zeta_3-6(1-z)\zeta_2+\frac{8}{27}(z+100)\Bigg]
+C_F T_F n_f\Bigg[\frac{224}{27}\frac{1}{(1-z)_+}+\frac{4}{9}\frac{1+z^2}{(1-z)_+}\nn\\
&(3H_{0,0}+5H_0)-\frac{4}{27}(19z+37)\Bigg]
+C_F^2\Bigg[2\frac{1+z^2}{(1-z)_+}(4H_{1,2}+4H_{2,0}+2H_{2,1}-2H_{1,0,0}\nn\\
&+4H_{1,1,0}+12\zeta_3)-\frac{2(2z^2-2z-3)}{1-z}H_{0,0}+12(1-z)H_{1,0}+\frac{2(16z^2-13z+5)}{1-z}H_0\nn\\
&+2(z+1)H_{0,0,0}+4(1-z)H_2+2z H_1+8(1-z)\zeta_2-22(1-z)\Bigg]\,.\nn
\end{align}
The soft function up to two loops is given by \cite{Li:2016ctv}
\begin{align}
\mathcal{S}_{\perp}\left(b,\mu,\nu\right)&=\exp \Bigg\{
\frac{\alpha_s}{4\pi}\left[ c_1^\perp+\frac{\gamma_{0}^{\rm cusp}}{2} C_F L_b^2+\gamma_{0}^{r} L_r
-L_b(\gamma_{0}^{s}+\gamma_{0}^{\rm cusp} C_F L_r)\right]\\
&+\left(\frac{\alpha_s}{4\pi}\right)^2\Bigg[
c_2^\perp+\gamma_{1}^{r}L_r+\frac{\gamma_{0}^{\rm cusp}}{6}C_F L_b^3 \beta_0
+L_b^2\left(\frac{\gamma_{1}^{\rm cusp}}{2}C_F-\frac{\gamma_{0}^{s}\beta_0}{2}-\frac{\gamma_{0}^{\rm cusp}}{2}C_F L_r \beta_0\right)\nn\\
&+L_b(-\gamma_{1}^{s}+c_1^\perp \beta_0+L_r(-\gamma_{1}^{\rm cusp}C_F+\gamma_{0}^{r}\beta_0))\Bigg]
\Bigg\}     \,, \nn
\end{align}
with
\begin{align}
 c_1^\perp&= -\frac{\pi^2}{3}C_F\,,\\
 c_2^\perp&=C_F C_A\left(-\frac{67\pi^2}{18}-\frac{154\zeta_3}{9}+\frac{\pi^4}{9}+\frac{2428}{81}\right)
+C_F T_F n_f \left(\frac{10\pi^2}{9}+\frac{56\zeta_3}{9}-\frac{656}{81}\right)\,,\nn\\
     \gamma_{0}^s &=  \, 0 \,,\nn \\
    \gamma_{1}^s &=  \, C_F C_A \left(\frac{11\pi^2 }{9}+28 \zeta_3-\frac{808}{27}\right)+ C_F T_F n_f \left(\frac{224}{27}-\frac{4 \pi^2}{9}\right) \,, \nn
\end{align}
and $L_r=\ln\left(\nu^2 b^2/b_0^2\right)$.  The expression for the subtracted jet function is expressed as \cite{Gutierrez-Reyes:2019vbx}
\begin{align}\label{eq:loop-jet-sub}
     J_q\left(b,\mu,\zeta\right) &= 1 +\frac{\alpha_s C_F}{4\pi}\left[-L_b^2 + L_b(3+2L_{\zeta})  -\frac{5\pi^2}{6}+7-6\,\ln2 \right]\\
 &+\left(\frac{\alpha_s}{4\pi}\right)^2  \Bigg\{
 C_F^2\Bigg[\frac{L_b^4}{2}-L_b^3(3+2L_\zeta)+L_b^2\left(2L_\zeta^2+6L_\zeta-\frac{5}{2}+6\ln 2+\frac{5\pi^2}{6}\right)\nn\\
 &+L_b\left(L_{\zeta}\left(14-12\ln 2-\frac{5\pi^2}{3}\right)+\frac{45}{2}
 -18\ln 2-\frac{9\pi^2}{2}+24\zeta_3\right)\Bigg]\nn\\
 &+C_F C_A\Bigg[-\frac{22}{9}L_b^3+L_b^2\left(\frac{11}{3}L_\zeta-\frac{35}{18}+\frac{\pi^2}{3}\right)+L_{\zeta}\left(\frac{404}{27}-14\zeta_3\right)\nn\\
 &+L_b\left(L_{\zeta}\left(\frac{134}{9}-\frac{2\pi^2}{3}\right)+\frac{57}{2}
 -22\ln 2-\frac{11\pi^2}{9}-12\zeta_3\right)\Bigg]\nn\\
 &+C_F T_F n_f\Bigg[\frac{8}{9}L_b^3+L_b^2\left(\frac{2}{9}-\frac{4}{3}L_\zeta\right)
 +L_b\left(-\frac{40}{9}L_\zeta-10+8\ln 2+\frac{4\pi^2}{9}\right)\nn\\
&-\frac{112}{27}L_\zeta \Bigg]
+j_2 \Bigg\}   \,, \nn
\end{align}
with $L_{\zeta} = \ln\left(\mu^2/\zeta\right)$.
\bibliography{refs.bib}

\bibliographystyle{JHEP}

\end{document}